\documentclass[useAMS,usenatbib]{mn2e}
\usepackage{epsfig}
\usepackage{epstopdf}
\newcommand\aj{{AJ}}%   }}+
\newcommand\araa{{ARA\&A}}%   
\newcommand\apj{{ApJ}}%   
\newcommand\apjl{{ApJ}}%   
\newcommand\apjs{{ApJS}}%   
\newcommand\apss{{Ap\&SS}}%   
\newcommand\aap{{A\&A}}%   
\newcommand\aaps{{A\&AS}}%   
\newcommand\mnras{{MNRAS}}%   
\newcommand\pasp{{PASP}}%   
\newcommand\sn{S$_{\rm N}$}          
\newcommand\zsun{{Z$_\odot$}}

\newcommand\msun{\rm{M$_{\odot}$}}

\newcommand\alfafe{[$\alpha$/Fe]}

\newcommand\teff{T$_{\rm eff}$}

\newcommand\rell{r$_{\rm ell}$}

\newcommand\sgperc{$\rm N_{\rm SG}/{\rm N}_{\rm tot}$ }
\def\simgt{\lower.5ex\hbox{$\; \buildrel > \over \sim \;$}}
\def\simlt{\lower.5ex\hbox{$\; \buildrel < \over \sim \;$}}
\title[]{The puzzle of metallicity and multiple stellar populations in the Globular Clusters in Fornax }
\author[F. D'Antona,  V. Caloi,  A. D'Ercole, M. Tailo, E. Vesperini, P. Ventura \& M. Di Criscienzo]{
F. D'Antona$^{1}$, V.Caloi$^{2}$,  A.D'Ercole$^{3}$, M. Tailo$^{4,1}$, E. Vesperini$^{5}$, P. Ventura$^{1}$, \newauthor  \& M. Di Criscienzo$^{6,1}$
\thanks{E-mail: franca.dantona@gmail.com (FD)
}
\\
$^{1}$ INAF, Osservatorio Astronomico di Roma, Via Frascati 33, 
I-00040 Monteporzio Catone (Roma), Italy.\\
$^{2}$ INAF, IAPS, Roma, via Fosso del Cavaliere 100, I-00133 Roma, Italy\\
$^{3}$ INAF- Osservatorio Astronomico di Bologna, via Ranzani 1, I-40127 BOLOGNA (Italy)\\
$^{4}$ Department of Physics, Universit\`a di Roma ``La Sapienza'', Roma, Italy\\
$^{5}$ Department of Astronomy, Indiana University, Swain West, 727 E. 3rd Street, IN 47405 Bloomington (USA)\\
$^{6}$  INAF, Osservatorio Astronomico di Capodimonte, Via Moiarello 16, I-80131 Napoli (Italy)\\
}
\begin{document}

\date{Accepted ... Received ...; in original form ...}

\pagerange{\pageref{firstpage}--\pageref{lastpage}} \pubyear{2009}

\maketitle

\label{firstpage}

\begin{abstract}
All models for the formation of multiple populations in globular
clusters (GCs) imply an initial mass of the systems several times
larger than the present mass.

A recent study of the dwarf spheroidal galaxy Fornax, where the low
metallicity ([Fe/H]$\simlt$--2) stars  contained in GCs appear to be
$\sim$20\% of the total, seems to constrain the initial mass of the
four low metallicity GCs in Fornax to be at most 
a factor of 5-6 larger than their present mass.

We examine the
photometric data for Fornax clusters, focussing our attention on their
horizontal  branch (HB) color distribution and, when available, on the RR
Lyr variables fraction and period distribution. Based on our
understanding of the HB morphology in terms of varying helium content (and
red giants mass loss rate) in the context of  multiple stellar generations, we show
that clusters F2, F3 and F5 must contain substantial fractions of
second generation stars ($\sim$54 -- 65\%). On the basis of a simple
chemical evolution model we show that the helium distribution in these
clusters can be reproduced by models with cluster
initial masses  ranging from values equal to $\sim$4 to $\sim$10 times
larger than the current masses. Models with a very short second generation star formation episode can also reproduce the observed helium distribution but require larger initial masses up to about twenty times the current mass.

While the lower limit of this range of possible initial GC masses is consistent with those suggested by the observations of the low metallicity field stars, we also discuss the possibility that the metallicity scale of field stars (based on CaII triplet spectroscopy) and the metallicities derived for the clusters in Fornax may not be
consistent with each other. In this case, observational constraints would allow larger initial cluster masses.

The reproduction of the HB morphology in F2, F3, F5 requires two
interesting hypotheses: 1) the first generation HB stars lie all at
``red" colours: namely, they populate only the RR~Lyr and the red
HB region. According to this interpretation, the low metallicity stars in the
field of Fornax, populating the HB at colours bluer than the blue side
---(V--I)$_0\simlt$0.3 or (B--V)$_0\simlt$0.2--- of the RR~Lyrs,
should be second generation stars born in the clusters;  a preliminary
analysis of available colour surveys of Fornax field provides a
fraction $\sim$20\% of blue HB stars, in the low metallicity range; 
2) the mass loss from individual second generation red giants is a few percent of a solar mass larger than the mass loss from first generation stars.
\end{abstract}

\begin{keywords}
stars: horizontal branch; stars: mass--loss; stars: evolution; globular clusters: general; galaxies: dwarf
\end{keywords}

\section{Introduction} 
 \label{sec:introduction}
The Fornax dwarf spheroidal galaxy (dSph) harbours 5 globular clusters
(GCs) \citep{hodge1961}, having masses, color magnitude diagrams and
ages similar to those of Galactic GCs \citep{buonanno1998}. High
dispersion spectroscopy of individual stars in these clusters
\citep{letarte2006} have recently shown that their stars
have abundance anomalies (oxygen, sodium and 
magnesium spreads) resembling those found in Galactic GCs. These
anomalies are the same 
that provided the initial evidence of multiple
stellar populations in Galactic GCs.

Fornax has a  high GC specific frequency \sn, defined as the number of GCs normalized to a host galaxy having M$_{\rm V}$ =--15~mag \citep{hvdb1981}: \sn=26 \citep{larsen2012a}, to be compared with the value \sn$\simeq$1 typical for spiral galaxies. This Òspecific frequency problemÓ is even more evident if related to the different stellar populations in galaxies, as \sn\ tends to increase with decreasing metallicity within galaxies (Harris \& Harris 2002; Harris et al. 2007), and this holds also for our own Galaxy.
The large fraction of Galactic GCs belonging to the halo constitutes a few percent of the halo mass, that indeed, at least in part, could be accounted for by disruption of GCs and loss of stars from them over a Hubble time. 
The problem of which fraction of the Galactic halo is made up by former GC stars has been emphasized again very recently, in the light of the models for the formation of multiple populations in GCs \citep{vesperini2010,schaerer2011}. 
Evidence is accumulating that the majority of stars in GCs belongs to the ``second generation" (SG), that turns out to be $>$50--80\% of stars for all clusters examined \citep{carretta2009a, dantonacaloi2008}.  
The main signature of this population, typical of GCs only \citep{gratton2012rev}, is the Na-O anticorrelation. The small fraction ($\sim$2.5\%) of Na-rich and/or CN-rich stars in the field halo (Carretta et al. 2010a; Martell \& Grebel 2010) may well come from SG stars evaporated from GCs.
All the models for the formation of the SG (both in the asymptotic--giant branch ---AGB--- scenario and in the rapidly rotating massive stars scenario) need that a great fraction of the first generation (FG) stars is lost from the clusters, in order to account for the present high percentage of SG stars. According to the specific choices for the parameters involved in the models, the ratio between initial mass of the FG and the present mass of GCs harbouring multiple populations must be in the range $\approx$5--20 \citep{bekki2007, dercole2008, decressin2007, renzini2008, carretta2009a, bekki2011}, and even up to 100 \citep{renzini2013}. 

The dynamical model by \cite{dercole2008}, based on the formation of a SG from
the ejecta of massive AGB stars and super--AGB stars, is taken as a basis by
\cite{vesperini2010} to determine the relationship between the fraction of cluster SG stars that are now in the halo and the general contribution of GC stars to the halo. Specifically, Vesperini et al. (2010)  showed that the observed fraction of SG stars in the halo (2.5 \%) implies that a fraction of about 20\%--40\% of the halo is composed of stars formed in GCs (if a Kroupa 2001 IMF is assumed; the range is 30 \% -- 60 \% if a Kroupa et al. 1993 IMF is instead adopted; see Vesperini et al. 2010 for further details).

Therefore, models for the formation and
evolution of multiple populations can shed light on the possible
contribution of GCs to the halo assembly and seem to suggest that
stars originally formed in GCs are a significant fraction of the
Galactic halo.  
Even more so, exploring the connection between multiple populations and the field
population of dSph galaxies with a high specific frequency of GCs can
provide additional insight and constraints on the formation and
dynamical history of GCs. \cite{larsen2012a} compared the
low-metallicity GC populations in Fornax with the corresponding field
star population  and concluded that the initial mass of these clusters can
not have been larger than a factor $\sim$5 their present mass. Moreover,
in the same range of metallicity, there can not have been other
clusters now dissolved in the field.\par 

In this paper  we estimate the fraction and the distribution of helium content of SG stars in the low metallicity GCs in Fornax following the analysis of the  HB morphology described   by \cite{dantona2002}. On this basis, we present a set of simple chemical models aimed at constraining the range of initial masses of GCs. We review the possible problems in the metallicity scales of GCs and field stars, and summarize how these uncertainties might affect the observational constraints on the GC initial mass.

%In this paper  we focus our attention on the distribution of stars along the HB of the low-metallicity GCs of Fornax dSph, and, following the interpretation of the HB morphology described   by \cite{dantona2002}, we estimate the fraction of SG stars in these clusters (Section \ref{sg}), with the help of models described in Section \ref{input}. 
%(Section \ref{chemevol}). We revise the possible problems in the metallicity scales of GCs and field stars (Section \ref{metals}), and finally summarize that these uncertainties might affect the observational constraints on the GC initial mass (Section 6).

\section{Models for the formation of Fornax clusters}
\label{input}

\subsection{Stellar evolution and synthetic horizontal branch computation}
We computed one basic set of isochrones and HB evolutions for this work. We adopt the metallicity Z=0.0003, and consider three helium contents, in mass fraction Y=0.25, 0.28 and 0.35. For an $\alpha$-enhancement \alfafe=0.4, with a choice of \zsun=0.018, this corresponds to [Fe/H]$\sim$--2, appropriate for several low metallicity Galactic GCs, although formally too large to describe the clusters F1, F2, F3 and F5 in Fornax (but see Section \ref{metals}). 
For T$>$10000 K we used the OPAL opacities, in the version documented by 
\cite{iglesias1996}, with all the recent updates; at lower temperatures the 
opacities by \cite{ferguson2005} were adopted. Conductive opacities were taken from the WEB site of
Potekhin (year 2006) \footnote{see http://www.ioffe.rssi.ru/astro/conduct}, corresponding to \cite{potekhin1999} treatment, corrected following the improvement of the treatment of the e--e scattering contribution described in \cite{cassisi2007}. 
Other inputs (convection model, non instantaneous mixing, treatment of the borders of convection) are all as described e.g. in \cite{dicriscienzo2011}.
We followed the main sequence and red giant (RG) evolution of low mass tracks. We compute isochrones from 9 to 14 Gyr for the chosen chemistry and the helium core mass at the helium flash. The RG evolution was followed at constant mass. The core masses are taken as input for the computation of HB models; the small helium increase in the envelope during the evolution is also taken into account.

Synthetic models for the HB are computed according to the recipes described in \cite{dantonacaloi2008}.
We adopt the appropriate relation between the mass of the evolving giant $M_{RG}$ and the age, as function of Y. 
The mass on the HB, as assumed in our previous work of HB simulations, is then simply: 
\begin{equation} 
M_{HB} =  M_{RG}(Y,Z) - \Delta M %\label{eq2} 
\end{equation} 
$\Delta M$\ is the mass lost during the RG phase. We assume that $\Delta M$\  has a gaussian dispersion $\sigma$\ around an average value $\Delta M_0$\  and  that both $\Delta M_0$\  and $\sigma$\ are parameters to be determined and {\it in principle} do not depend on Y.  We will see that a reasonable fit of the morphology of the Fornax clusters HBs requires that {\it the mass loss of the second generation stars is larger than in the FG stars}, by a small but significant amount. We then modify the expression to:
\begin{equation} 
M_{HB} =  M_{RG}(Y,Z) - \Delta M - \delta M_{SG}%\label{eq2} 
\end{equation} 
\noindent
so that $\Delta M$\ is fixed for both populations in order to fit the
redder HB stars that we attribute to the FG, and $\delta M_{SG}$\ will
be fixed, along with a number vs. helium distribution assumed for
the SG, in order to satisfactorily reproduce the bluer parts of the
HB.  For given values of Z and Y, the \teff\ location of an HB mass is fixed. Consequently, different ages can be adopted, provided that the mass loss is consistently adjusted.  The RR Lyr variables are identified as those stars that, in the simulation, belong to the \teff\ interval $3.795 < \log T_{\rm eff} < 3.86$. Their periods are computed as in  \citet[][see their Eq. 1]{dicriscienzo2004}.

\subsection{Chemical evolution models}
\label{sec:model}
Once acquired the helium distribution function N(Y) that best accounts
for the star distribution along the cluster HB, we attempt to
reproduce it by means of \cite{dercole2010,dercole2012} chemical
evolution model, hereafter referred to as D2010 and D2012.  According
to the framework presented in D2010, FG stars are already in place and
have the same chemical abundances of the pristine gas from which they
form; the SG stars form from super--AGB and AGB ejecta collecting in
the cluster centre through a cooling flow, partially diluted with
pristine gas also accreting in the cluster core. For the Fornax
clusters, we have adopted the yields from the latest computations by
\cite{ventura2013}, that refer to the global metallicity Z=0.0003 (as
specified above). 
There are not enough spectroscopic data to test the results of our
models on other 
elements abundances (e.g. oxygen and sodium abundances), and here we
focus only on the Y distribution, $N(Y)$. 
The chemical evolution models consistent with the $N(Y)$ derived from
the HB observations will provide the ratio of the SG to the clusters initial masses (see Section 5).
\begin{figure*}    
\centering{
\includegraphics[width=5cm]{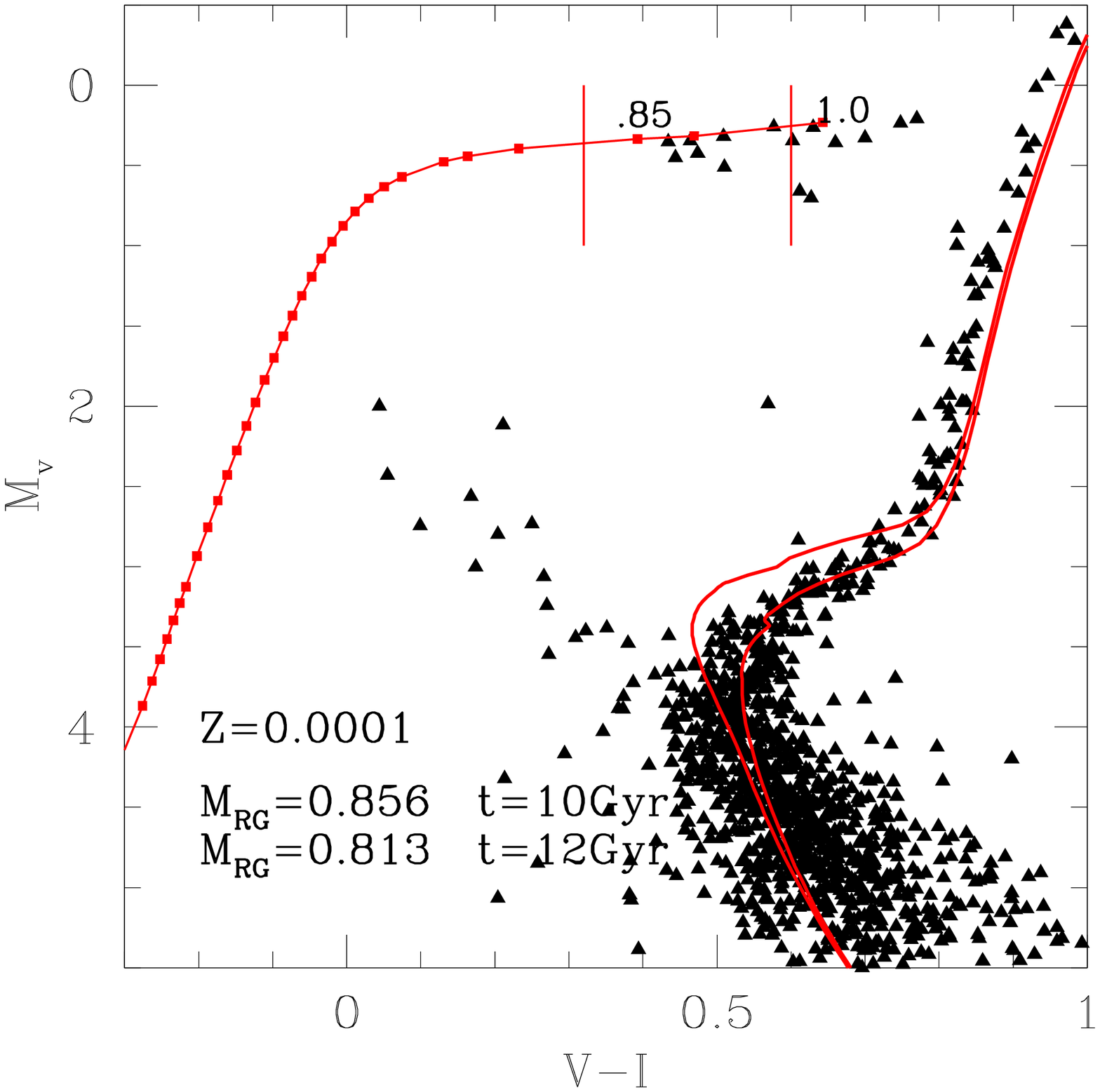}
\includegraphics[width=5cm]{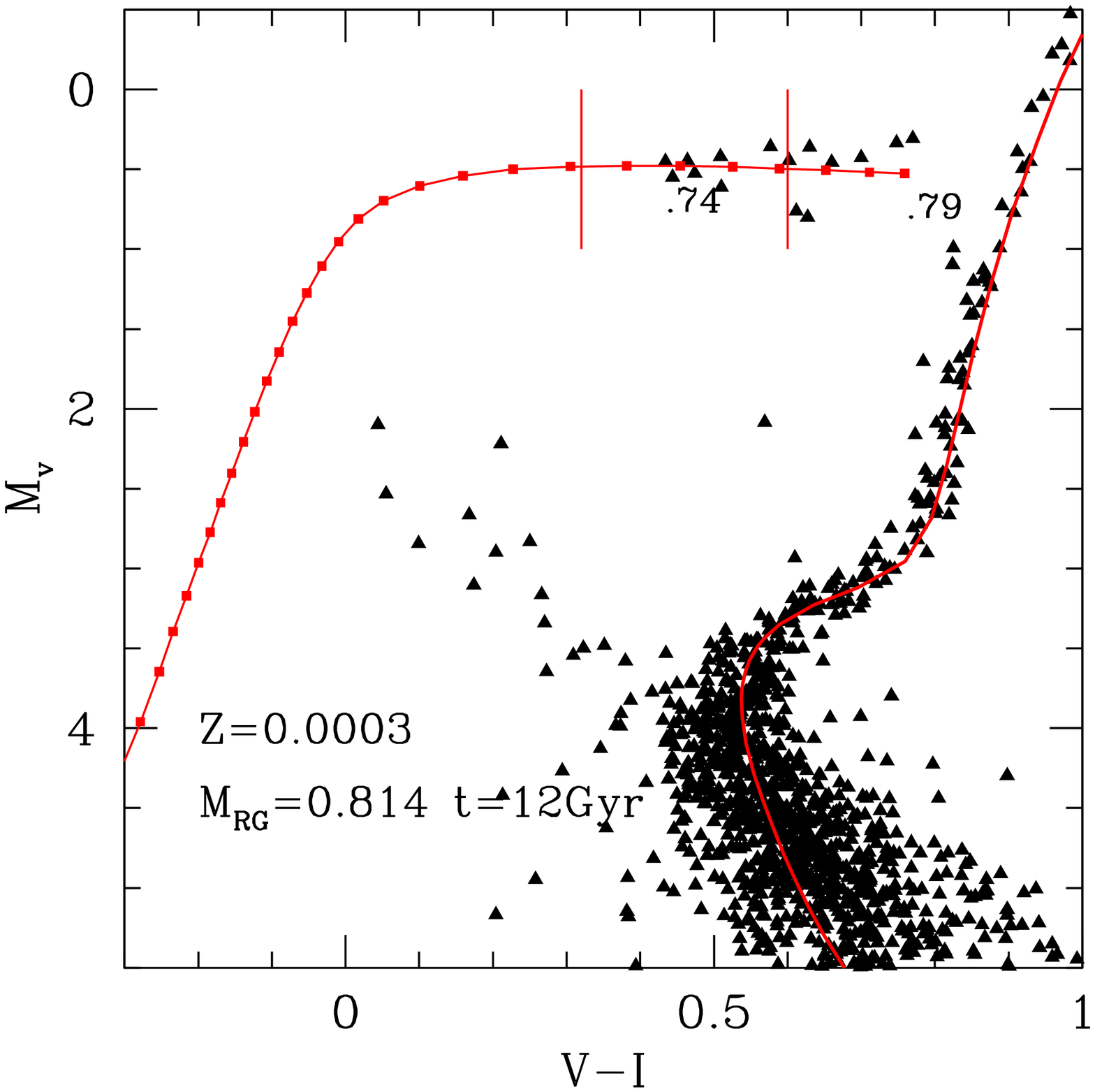}
\includegraphics[width=5cm]{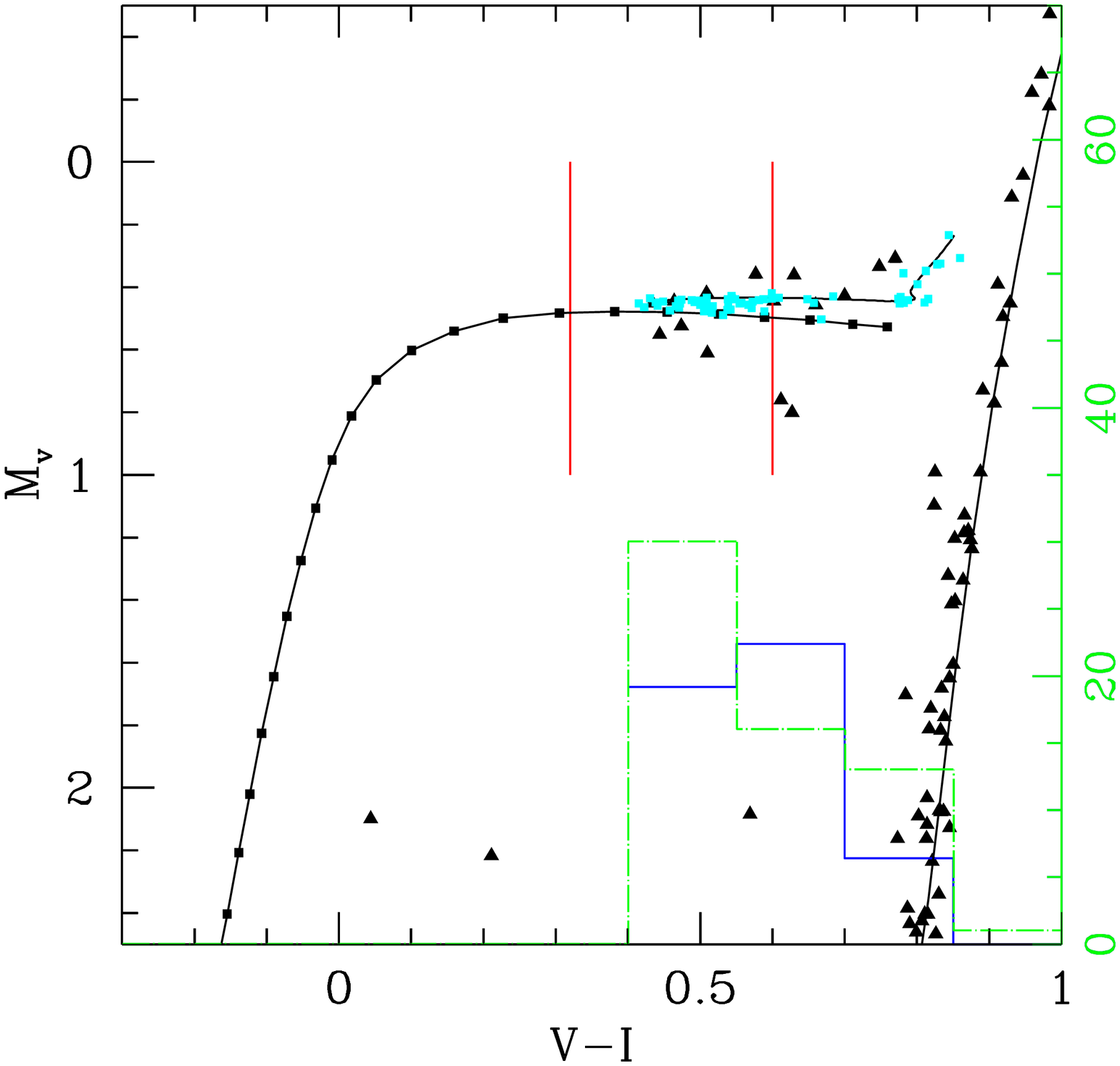}
}
\caption{The color magnitude diagram of cluster F1 from Buonanno et al. 1998 is shown. In the left panel, the data have been reported to absolute magnitude and intrinsic colour by assuming a reddening E(V-I)=0.02 and a distance modulus (V-M$_{\rm v}$)=20.85; the choice has been made to allow for a match with the ZAHB location of models with Z=10$^{-4}$. Along the ZAHB, the positions of 0.85 and 1.0 \msun\ are indicated; the isochrones and corresponding RGB mass for ages 10 and 12Gyr are given.  We see that an age of 10Gyr is barely compatible with the HB data, while the relative morphology of HB and turnoff are not.
In the central panel, we show the same data, assuming the same dereddening and  (V-M$_{\rm v}$)=20.75, to compare with ZAHB and isochrones for Z=3$\times10^{-4}$. We see that for this Z there is ample space in the mass loss in order to reproduce the HB. In the right panel, we show a simple simulation of the HB data, obtained with a unique population of FG stars at Y=0.25, Z=3$\times 10^{-4}$. At the bottom, we show the histogram of the number of HB stars versus V--I (full line) compared with the histogram of the simulation (dashed).
} 
\label{f1} 
\end{figure*}

\section{The metallicity of GCs and field stars}
\label{metals}

\subsection{The metal--poor component in the Fornax field and CaII triplet  calibration}

In the last twenty years CaII triplet spectroscopy applied to GC red
giants has been frequently used as a proxy for Fe--abundance
\citep[e.g.][]{dca1995,  rutledge1997, dch1998}. The method has been
extended  to intermediate age clusters and field stars of various age
and composition in nearby Galactic satellites
\citep[e.g.,][]{tolstoy2001,cole2004}.  The advantages and the
drawbacks of the method are discussed in detail in  those papers and references therein. As pointed out, among others, by \cite{dch1998}, the theoretical basis of the method is not well understood, since the dependence of the intensity of the CaII triplet on Ca-- and Fe-- abundances is not known. Besides, the method is calibrated on GC giants, where [Ca/Fe] is assumed to be about +0.3; if used for giants with lower values of [Ca/Fe], it will lead to an underestimate of [Fe/H]. Another important point is the choice of the abundance scale for GCs, which of course will influence the Ca--triplet calibration.

In the case of Fornax field, the situation is discussed in detail by \cite{tolstoy2001}, \cite{battaglia2006, battaglia2008} and \cite{starkenburg2010}. Tolstoy et al. note that  the abundances of individual stars do not always correspond to their    colour, an occurrence to be kept in mind when considering the possible  loopholes of the method.  At the same time, \cite{battaglia2008} and \cite{starkenburg2010} propose the ultimate solution of the problem, with a direct calibration of the Ca--triplet through high resolution spectroscopy of 36 giants in the Fornax field. The latter paper extends the relation found in the  former down to a very low Fe--content [Fe/H] $\simeq$ --4 dex. In this  Fe--abundance range the errors increase, but not to the point of affecting the estimate of an extreme Fe--deficiency (Starkenburg et al. 2010, Fig. 10).

As a matter of fact, the sample of high resolution observations in  \cite{battaglia2008} consists of 36 RG stars, for 33 of which the estimated  [Fe/H] is $\ge -1.0$, while for the remaining three we have [Fe/H] =  --1.44, --1.53 and --2.65. 
Besides, the estimates for these three stars through the adopted CaII calibration, based on Carretta  \& Gratton scale (1997), are [Fe/H] = --1.81, --1.85, --2.39, values significatively different from those derived through high resolution spectroscopy. Therefore we consider difficult to establish a reliable calibration of the low metallicity tail in Fornax field on this basis.
There is another important point: recently, the Carretta \& Gratton scale
   has been substantially changed \citep{carretta2009iron}. According  to the
   authors, among the main reasons behind this result are: the adopted
   temperature scale and the difference in the {\it gf}--values. In any case,
   the CaII triplet calibrations should be reconsidered. We observe also that
   the extreme sensitivity of high resolution spectroscopy on the  adopted parameters for spectra interpretation is still not under control and should suggest caution in the estimates of the errors 
involved. 
 
 An interesting point is raised by \cite{starkenburg2010}:  Fornax exhibits a low--metallicity tail much less pronounced than in many other dwarf speroidals and ultra--faint galaxies (see their Fig.14). They wonder whether this fact is due to a different chemical history in the early epochs, or if the low metallicity population is being   hidden by the dominant young and metal--rich population (this would be possible since the observed sample in Fornax is a very small fraction of the total number of RG stars in this galaxy). 

\subsection{The determination of [Fe/H] in Fornax clusters}

The precise metallicity of Fornax clusters is a key element for our analysis.
Strader et al. (2003) collect and discuss their [Fe/H] estimates along with
those preceding their work. These estimates are based on low resolution
spectra, various photometric indices and CM diagram features, and give values
never lower than --2.1 dex.  Recently, high resolution spectroscopy of both
individual stars and integrated cluster light have given much lower values of
[Fe/H] for F1, F2, F3 and F5 .  For example, \cite{larsen2012b} obtain [Fe/H]
= --2.35 for F3 through the analysis of integrated cluster spectra. The
spectra are interpreted by means of synthetic population analysis based on
stellar counts \citep[from the photometric data by][]{buonanno1998},
extrapolated to lower luminosity by appropriate isochrones
\citep{dotter2007}. They say that this value agrees ``reassuringly well" with
the value of [Fe/H] = --2.4$\pm$0.1 derived by Letarte et al.  (2006) from
high dispersion spectroscopy of three stars in F3. As mentioned before, this
is significantly lower than both the old value listed by \cite{buonanno1998}
of --1.96$ \pm$0.20, based on the slope of the RG branch, and the more recent
one of --1.84$\pm$0.18 given by \cite{strader2003} on the basis of
low--resolution integrated spectra. Another estimate comes from
\cite{greco2007}, who find [Fe/H]=--1.91 from the mean period of
ab--type RR Lyrs, using Sandage (1993) relation.  Similar
considerations hold as well for F2, F5 and F1. For this latter cluster
observational data are scarcer, but \cite{letarte2006} find
[Fe/H]=--2.5$\pm$0.1 (again from high--dispersion spectra of three
members), significantly lower than the values listed by
\cite{strader2003}. 

Although a review and analysis of  existing observations is well
beyond the scope of this work, we will argue in the following sections 
that values of [Fe/H] lower
than $\sim$--2.2 can be excluded on evolutionary considerations.

\subsection{The conversion between [Fe/H] and metals mass fraction}

The global metallicity, that is the metal mass fraction Z which enters
in stellar structure computation, can be derived by choosing the
elemental solar ratios of abundances, an appropriate relation between
metallicity, [Fe/H] and \alfafe, and the solar metal abundance. We
derived the following two relations: the first one based on the
\cite{gs98} (GS98) elemental solar ratios.

\begin{table*}
\begin{center}
\caption{Conversion [Fe/H] -- Z}
\begin{tabular}{c cccccc}
\hline\hline
[Fe/H]   &   \multicolumn{3}{c}{Z$_\odot$=0.014  AS09 }      &    \multicolumn{3}{c}{Z$_\odot$=0.018  GS98 } \\      
             &   $[\alpha$/Fe]=0.1 &  $[\alpha$/Fe]=0.2 & $[\alpha$/Fe]=0.4 &  $[\alpha$/Fe]=0.1 &  $[\alpha$/Fe]=0.2 & $[\alpha$/Fe]=0.4 \\
 --2.5     &    $5.2 \times 10^{-5}$    &     $6.0 \times 10^{-5}$   &    $7.9 \times 10^{-5}$   &  $6.7 \times 10^{-5}$   &  $7.8 \times 10^{-5}$   &  $9.1 \times 10^{-5}$  \\
 --2.4    &    $6.5 \times 10^{-5}$    &     $7.5 \times 10^{-5}$   &    $1.2 \times 10^{-4}$   &  $8.4 \times 10^{-5}$   &  $9.8 \times 10^{-5}$   &  $1.3 \times 10^{-4}$  \\
 --2.3    &    $8.2 \times 10^{-5}$    &     $9.5 \times 10^{-5}$   &    $9.9 \times 10^{-5}$   &  $1.1 \times 10^{-4}$   &  $1.2 \times 10^{-4}$   &  $1.7 \times 10^{-4}$  \\
 --2.2    &    $1.0 \times 10^{-4}$    &     $1.2 \times 10^{-4}$   &    $1.6 \times 10^{-4}$   &  $1.3 \times 10^{-4}$   &  $1.6 \times 10^{-4}$   &  $2.1 \times 10^{-4}$  \\
 --2.1    &    $1.3 \times 10^{-4}$    &     $1.5 \times 10^{-4}$   &    $2.0 \times 10^{-4}$   &  $1.7 \times 10^{-4}$   &  $2.0 \times 10^{-4}$   &  $2.6 \times 10^{-4}$  \\
 --2.0    &    $1.6 \times 10^{-4}$    &     $1.9 \times 10^{-4}$   &    $2.5 \times 10^{-4}$   &  $2.1 \times 10^{-4}$   &  $2.5 \times 10^{-4}$   &  $2.9 \times 10^{-4}$  \\
 --1.5    &    $5.2 \times 10^{-4}$    &     $6.0 \times 10^{-4}$   &    $7.9 \times 10^{-4}$   &  $6.7 \times 10^{-4}$   &  $7.8 \times 10^{-4}$   &  $1.0 \times 10^{-3}$  \\
 --1.4    &    $6.5 \times 10^{-4}$    &     $7.5 \times 10^{-4}$   &    $9.9 \times 10^{-4}$   &  $8.4 \times 10^{-4}$   &  $9.8 \times 10^{-4}$   &  $1.3 \times 10^{-3}$  \\

\hline
\end{tabular}
\label{table1}
\end{center}
\end{table*}

\begin{equation} 
\log\left( Z / Z_\odot \right) = [Fe/H] + 0.72 [\alpha/Fe] - 0.15  [\alpha/Fe]^2 + \log\left( X /X_\odot \right) 
\end{equation} 
\noindent
and the second one based on the elemental ratios by \cite{asplund2009araa} (AS09):

\begin{equation} 
\log\left( Z / Z_\odot \right) = [Fe/H] + 0.69 [\alpha/Fe] - 0.16  [\alpha/Fe]^2 + \log\left( X /X_\odot \right) 
\end{equation} 

The global metallicity Z is obviously dependent also on the solar metallicity Z$_\odot$. The value (Z/X)$_\odot\simeq$0.023  applies to \cite{gs98} determination, and a value  (Z/X)$_\odot \simeq$0.018\footnote{Notice that the ratio (Z/X)$_\odot$\ is the value observed at the solar photosphere. The action of diffusion in the 4.6 billion years of solar lifetime is such that the hydrogen mass fraction is increased to X$_\odot\simeq$\ 0.76, so that the resulting solar mass fractions are  \zsun$\simeq$0.018 for GS98, and \zsun$\simeq$0.014 for AS09. } corresponds to \cite{asplund2009araa} determination. The downward revision  is a result of the analysis of solar spectral lines based on 3D hydrodynamic models of the solar atmosphere, a careful selection of spectral lines, and relaxing the assumption of LTE. Nevertheless, we have to mention the ``solar abundance problem" emerged with the new solar abundances.
The \cite{gs98} value of (Z/X)$_\odot$ allowed the construction of a ``standard
solar model" perfectly matching the detailed structure of the solar
interior that emerged from helioseismic studies \citep{cd1996,
  bahcall2001}. On the contrary, computation of solar models incorporating reduced CNO abundances
\citep{bahcall2004, basu2004, montalban2004} showed that the location of the
boundary of the solar convective envelope, the sound speed and density
profiles, and today's surface helium abundance predicted by models using low
(Z/X)$_\odot$, differ from the values obtained from helioseismology by amounts
beyond accepted uncertainties \citep{bahcall2006}. Models in satisfactory
agreement with these data can be built, but they rely on parametric adjustments either of interior opacities, or of other pieces of input physics  \citep{bahcall2005, montalban2006}.

\subsection{Problems concerning the metallicity of globular clusters F1 and F4}
\label{dating}

The question we intend to pose appears very clear when discussing the cluster F1. We shall discuss the maximum possible range of metallicities that correspond to [Fe/H]=$-2.5 \pm$0.1 estimated for F1 by \cite{letarte2006}. For the sake of argument, let us assume \alfafe=0.2 \citep[see the estimates of calcium for Fornax clusters in][]{larsen2012b}. 

From the correspondence between [Fe/H] and metal mass fraction listed in Table \ref{table1}, we see that a value of Z=10$^{-4}$\ for \alfafe=0.2 overestimates the abundances measured by \cite{letarte2006} in individual cluster stars. We then used the isochrones and horizontal branch models computed for this composition by \cite{dicriscienzo2011} for comparison with F1 data.  The left panel of Fig. \ref{f1} shows that  an age of 12\ Gyr is incompatible with the very red HB of this cluster: in fact, the evolving RGB mass is a bare 0.81 \msun, while the blue edge of the HB requires a zero age HB value of 0.85 \msun. Even if no mass is lost along the
RGB, the cluster can not be older than $\sim$9.5 Gyr. Such a value is in contrast
with the relative location of the HB and the turnoff and, besides,
\cite{buonanno1998} showed that the loci of the four clusters F1, F2, F3 and
F5 can be superimposed on each other, as well as  on the old metal poor GCs
M~68 and M~92. 

In order to obtain a reasonable fit of F1 CM diagram, we move to isochrones and HB models of Z=0.0003 and \alfafe=0.4, that clearly overestimate the metallicity and \alfafe\ of this cluster.  In this case, the RGB mass is still 0.81 \msun\ at 12 Gyr, but the HB locations at fixed mass have moved to redder colors, thanks to the larger efficiency of the H--burning shell, and we find a value of 0.75 \msun\ for the bluest HB objects of F1 (Fig.\ref{f1}, central panel). Therefore, there is enough space between Z=0.0001 and Z=0.0003 to achieve a reasonable fit of the CM diagram features of this cluster. While core hydrogen burning and lifetimes do not strongly depend on the metallicity at such low values ---and the evolving mass does not change by shifting Z from 1 to 3 $\times 10^{-4}$--- the location in \teff\ of masses along the HB is in fact linked to the relative efficiency of He--core burning and H--shell burning in the star: by increasing the metallicity,  we
  increase the CNO in the hydrogen shell, and make the shell burning more
  efficient, so that smaller masses move to lower \teff. In order to locate a
  mass equal to the evolving RG mass (0.81\msun at 12Gyr) at the bluest
  location of F1 HB, it is necessary to increase Z to $\sim 1.5 \times
  10^{-4}$, that is to [Fe/H]$\sim -2.3$ to $-2.2$\ (according to the chosen
  Z$_\odot$).  
  
  The study of the exact composition of F1 stars is beyond the scope of this paper, we only wish to point out that the ``formal"  metallicity derived from high dispersion spectroscopy is not compatible with cluster features. As for the problem of the formation of  this cluster, we also can point out that looks like a ``first generation only" cluster, in the sense of \cite{caloi2011}. In fact, the distribution  of its HB stars with colour can be interpreted as due to the evolution of masses M=0.74$\pm 0.005$\msun, deriving from a simple  stellar population with cosmological Y content aged 12 Gyr that has  lost $\sim$0.07\msun\ on the RGB. The result is shown in the right panel of Fig.\ref{f1}, with the simulated HB population and the corresponding histogram (dash dotted). 

The small current mass of F1 might be consistent with the lack of a SG population (although it is important to point out that the cluster mass and structural properties which are relevant for the SG formation are the initial ones and those might significantly differ from the current ones).
In this context, we notice that \cite{letarte2006} provide abundances for three stars in F1, and find remarkable abundance differences in sodium, oxygen and magnesium. If confirmed, these are not consistent  with our interpretation, but the sample is too limited to draw any conclusion.

\begin{figure}    
\centering{
\includegraphics[width=8cm]{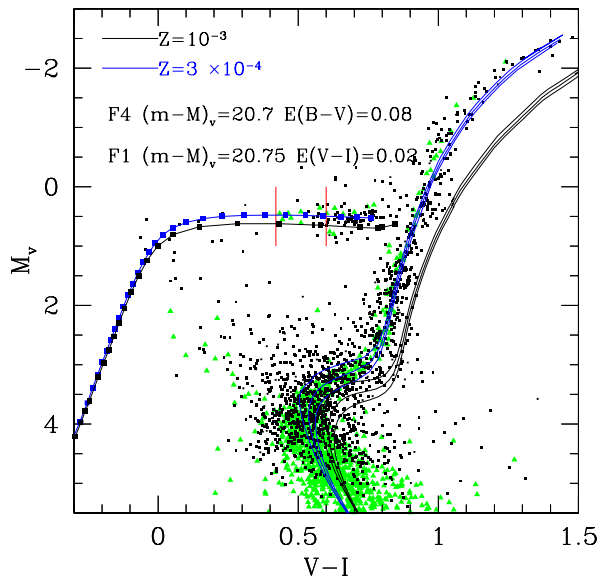}
}
\caption{The data for clusters F1 (green triangles, from Buonanno et al. 1998)
  and F4 (black dots, from Buonanno et al. 2009) are superimposed by assuming for F4 the minimum reddening that allows to populate the RR~Lyr strip (Greco et al. 2009) and a distance modulus compatible with the plotted ZAHB locations of the Z=0.0003 and Z=0.001 models. We also show isochrones of 10, 12 and 14~Gyr for both metallicities (the Z=0.001 isochrones are redder). As well documented in Bunonanno et al. 2009, we see that the morphology of the RGB and turnoff regions are not compatible with a difference in metallicity of a factor 10 between the two clusters.} 
\label{f2} 
\end{figure}

In this work, we leave aside the discussion of cluster F4; its role is not particularly relevant for the problem of multiple  population formation, as its most recent [Fe/H] value lies in the range where field stars are very abundant. Nevertheless, it is interesting to point out some issues concerning the metallicity of this cluster.

First of all, the range of [Fe/H] attributed to F4 goes from --1.4 to --2.2 (summary Table 5 in\cite{strader2003}). A value [Fe/H]=--1.4 is also found by \cite{larsen2012a}. The $\alpha$--enhancement, if any, is very low, so the global metallicity we can derive from Table \ref{table1} is 6--8$\times 10^{-4}$. The reddening values E(B--V)=0.08--0.12 quoted in the literature are mandatory to place at least the bluest part of its HB stars into the RR~Lyr gap (many RR~Lyr variables are present in the cluster \cite{greco2009}). Once applied the lowest reddening  E(B--V)=0.08mag, and a distance modulus consistent with the level of the HB, we re--discover the result largely discussed in \cite{buonanno1999}, namely, that the turnoff and RG colours result very similar to the colours of cluster F1 (Fig.~\ref{f2}). Also the RGB slopes are similar, a finding that led \cite{buonanno1999} to propose an [Fe/H] $\simeq$--2 for this cluster. We remark that it is difficult to believe in a metallicity difference of 1.1~dex (from --2.5 for F1 to --1.4 for F4) as proposed in the recent determinations. 

One can suggest that the information that the HB stars occupy only the RR~Lyr region and the red HB is in favor of a larger average metallicity, but we have seen that the HB of F1 is red too, and will discuss in Sect.~\ref{sg} that we expect not very blue HBs for the FG of the other clusters. 
As the HB of F4 is not very extended in color, we are led to interpret it as a FG--only population. This is however in contrast with the relatively large mass of the cluster, and especially with its concentration. Further effort is necessary to understand the precise role of star formation in this cluster.

\begin{figure*}     
\centering{
\includegraphics[width=8cm]{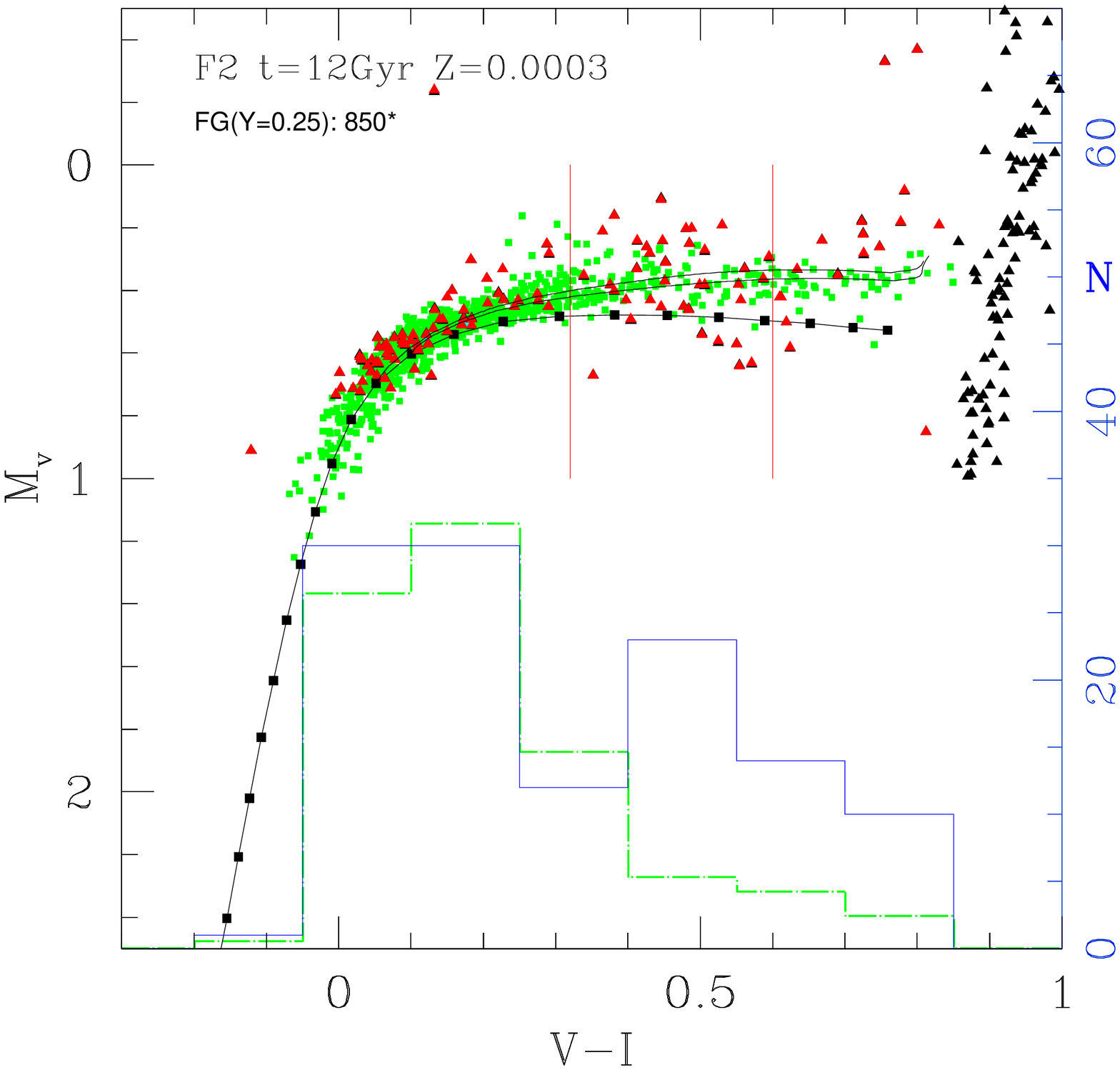}
\includegraphics[width=8cm]{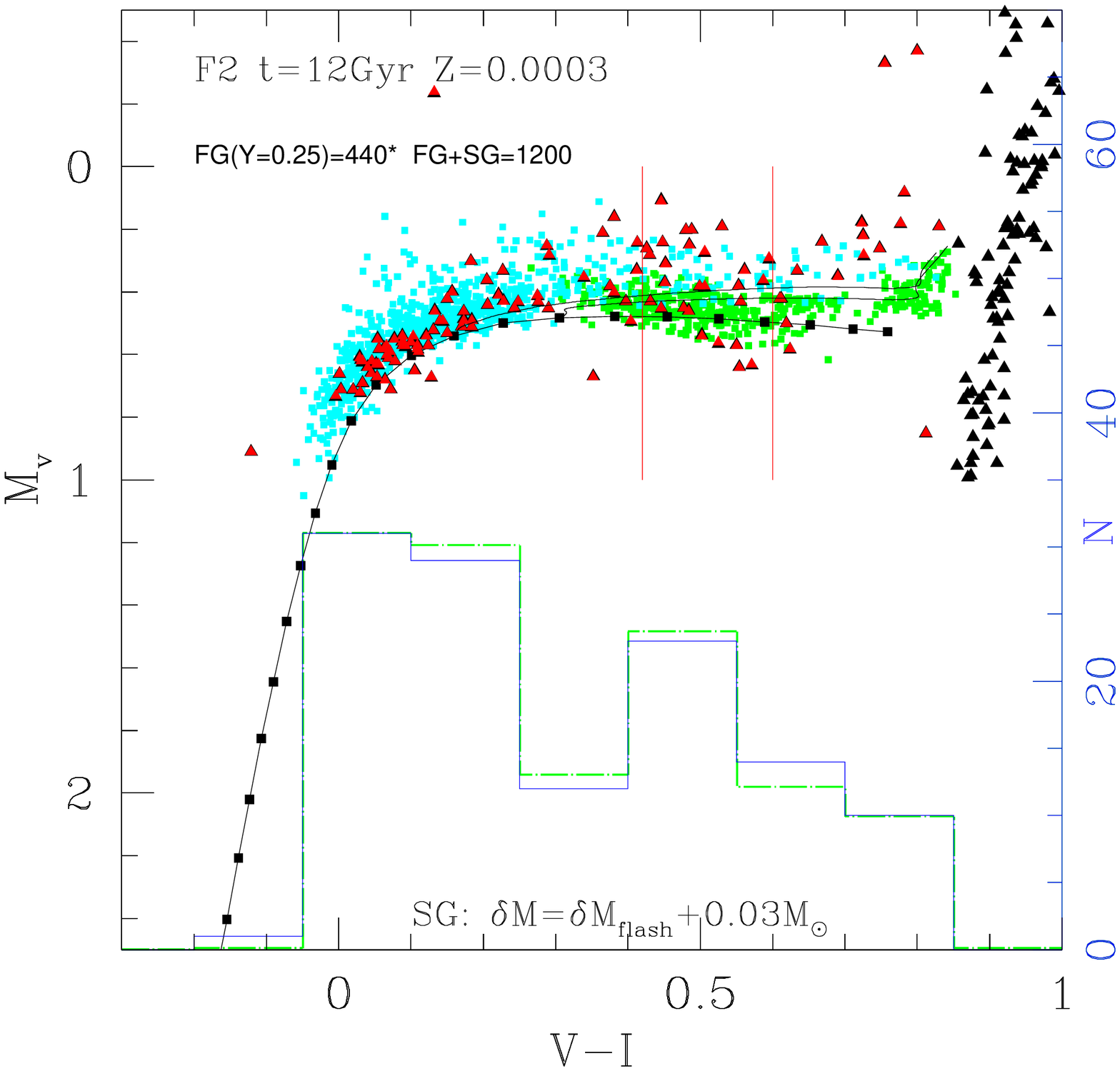}
}
\caption{The panels show the HST data in the bands F555 (labelled V) and F814
  (labelled I) for the cluster F2, from Buonanno et al. 1998 (red and black
  triangles). The full line connecting the black squares represents the ZAHB
  for the models of Z=0.0003 and Y=0.25. Two HB tracks are shown, for M=0.68
  and 0.69\msun\ in the left panel, for M=0.70 and 0.72\msun\ in the right
  panel. The RR~Lyr gap is approximately limited by the vertical red
  lines. Red triangles indicate the stars we consider as HB members (120), for which
  the full line histogram at the bottom shows the number vs. colour
  distribution. The simulations are shown by green squares (FG) and cyan
  squares (SG): see text for details. The corresponding colour distributions are shown in both panels as 
  dot--dashed histograms.
The left panel shows the HB data for F2, together with an attempt of
simulating them by means of a unique FG population (model~2 in
Table~\ref{table2}): RR~Lyrs and red HB stars are not accounted for. The right panel shows the comparison with simulation model 4, able to reproduce the peak of stars in the RR Lyr region. This assumes the presence of a SG population having larger helium abundances and larger  mass loss on the RGB, by 0.03\msun.} 
\label{f3} 
\end{figure*}

\begin{table*}
\begin{center}
\caption{Simulations parameters}
\begin{tabular}{c ccccccc}
\hline\hline
 & Cluster &   Z   & age(Gyr) &   $\delta$M$_{RG-HB}$/\msun      &      $\delta$M$_{RG add.}$/\msun  & $\sigma$(\msun) & 
  \sgperc \\  % &  FG/SG  \\
\hline
1 &F1       &   0.0003  &12            &   0.069            &              0                      &  0.012 & 0  \\
2 &F2       &  0.0003   &12            &   0.13         &              0                      &  0.015  &   0    \\
3 &F5       &  0.0003   &12            &   0.130            &              0                      &  0.015 & 40  \\
\hline
\\
4 &F2     &  0.0003 &12            &   0.0735              &              0.03           &  0.004  & 58  \\% & 0.71 \\
5 &F3       &  0.0003 &12            &   0.0713            &              0.03           &  0.009 &  54   \\% &  0.85 \\
6 &F5       & 0.0003 &12             &   0.0770             &              0.03           &  0.008 &  65   \\% & 0.54\\
\hline
\end{tabular}
\label{table2}
\end{center}
\end{table*}

\section{The extent of the SG in clusters F2, F3 and F5}
\label{sg}
At first sight, the CM diagrams of clusters F2, F3 and F5
\citep{buonanno1998} may be misleading. Cluster F2 has a ``short" blue HB, that could resemble the luminous part of the very extended HB of NGC 2419, interpreted by \cite{dicriscienzo2011} as the FG fraction of the cluster stars. The RR Lyrs in NGC 2419 are not on the zero age HB, but lie on tracks starting on the blue HB: however, their luminosity is close to the zero age luminosity (as appropriate for very metal poor evolutionary tracks). The HB in F5 looks like that in F2, but it has a short blue extension, so we could be led to think that this extension represents a small SG with a slightly increased helium content \citep[like in the cluster M~53, in the simulations by][]{caloi2011}. Only F3 has a very extended HB that resembles a cluster with a high percentage of SG. If this were the case, the problem of the small fraction of stars having the clusters metallicity in Fornax field would be simple to solve, as there is only one cluster, F3, with a second generation that requires a strong FG mass loss during the early phases of this cluster dynamical evolution.

A closer look at the HBs, however, shows that this is not the case,
and that all the three cluster have a large fraction of SG
stars. We examine them in order of increasing colour extent of their
HB. 

\subsection{Simulations of the HB in F2}
In F2, \cite{letarte2006} find [Fe/H]=--2.1$\pm$0.1 from high dispersion spectra. We use our evolutionary tracks for Z=0.0003 and standard helium content Y=0.25 to describe the FG of this cluster. The distance modulus and reddening are chosen, within reasonable values, to fit our own models. For this cluster we adopt (V--M$_{\rm v}$)=20.88 and E(V--I)=0.05. We use a ratio E(V--I)/E(B--V)=1.3, and an absorption A$_{\rm v}$=3.1E(B--V).
We first attempt  an ``FG only'' description of the HB. The left panel of
Fig. \ref{f3} shows the result. The age chosen for the cluster is a standard
12Gyr, the mass loss on the RGB had to be fixed at $\Delta$M=0.13\msun, to fit
the bluer end of the HB, and a mass dispersion $\sigma$=0.015\msun\ has been
chosen to spread the HB as required (model 2 in Table 2). The evolving masses
have a peak in the range 0.68--0.69\msun. Nevertheless, it is well clear that
the number of stars in the RR Lyr region {\sl where the colour distribution
  shows a secondary maximum} can not be reproduced, as the models spend most
of their lifetime in the blue, and evolve fastly at the RR Lyr location. The
total number of HB stars is 120, but  the total number of simulation stars
shown is 850, and the histogram plots the number in each bin divided by 10,
corresponding to a total of 85 stars. With such a choice, we  normalize the
simulation to the number counts in the bins at the left of the RR~Lyr
gap\footnote{If we normalize to the total number of observed stars (120), the
  comparison amplifies the inconsistency between the color distribution of the
  data set and that of the simulation, as most of the extra stars would raise the number at the blue HB side, while the red side would still show fewer stars than observed. }  From this comparison, we see that the star distribution along the HB is very different from the distribution in the luminous part of the HB of NGC 2419: in the latter cluster the peak in the blue is much more pronounced; besides, the RR~Lyr and red HB stars can not be ascribed to the evolved part of the HB tracks starting in the blue HB peak.

In order to fit the HB of this cluster, it is necessary to reduce the mass loss along the RGB for the FG, and to reproduce the peak in the RR Lyr gap {\it as the locus of evolution of the FG stars} \citep[see the simulations for M~3 in][]{caloi2005}. The successful simulation parameter are those of model 4 in Table 2.
In order to reproduce the bluer parts of the HB, we attribute them to a SG, with increasing values of the helium content for increasing \teff\ (right panel of Fig.\ref{f3}). To obtain a satisfactory fit to the observed HB morphology in the colour--magnitude diagram, we had also to add a $\delta$M$_{SG}$=0.03\msun\ to the mass loss of the SG. We will discuss the necessity of this choice in Sect.~\ref{F6}.
As a result, the FG population in this cluster, obtained from the HB morphology, is  42\% of the total.

\subsection{Simulations of the HB in F5}
\begin{figure*}    
\centering{
\includegraphics[width=5cm]{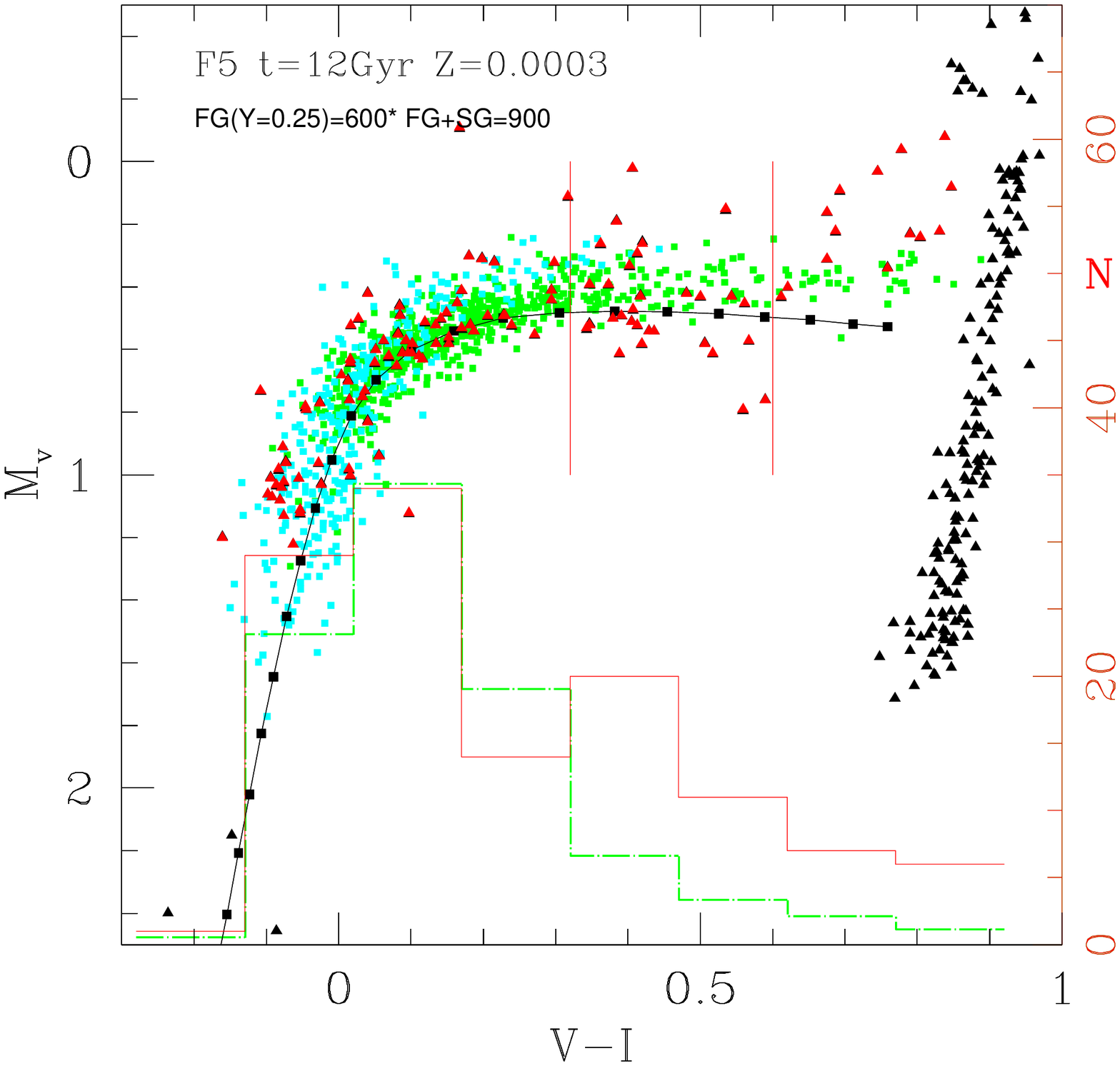}
\includegraphics[width=5cm]{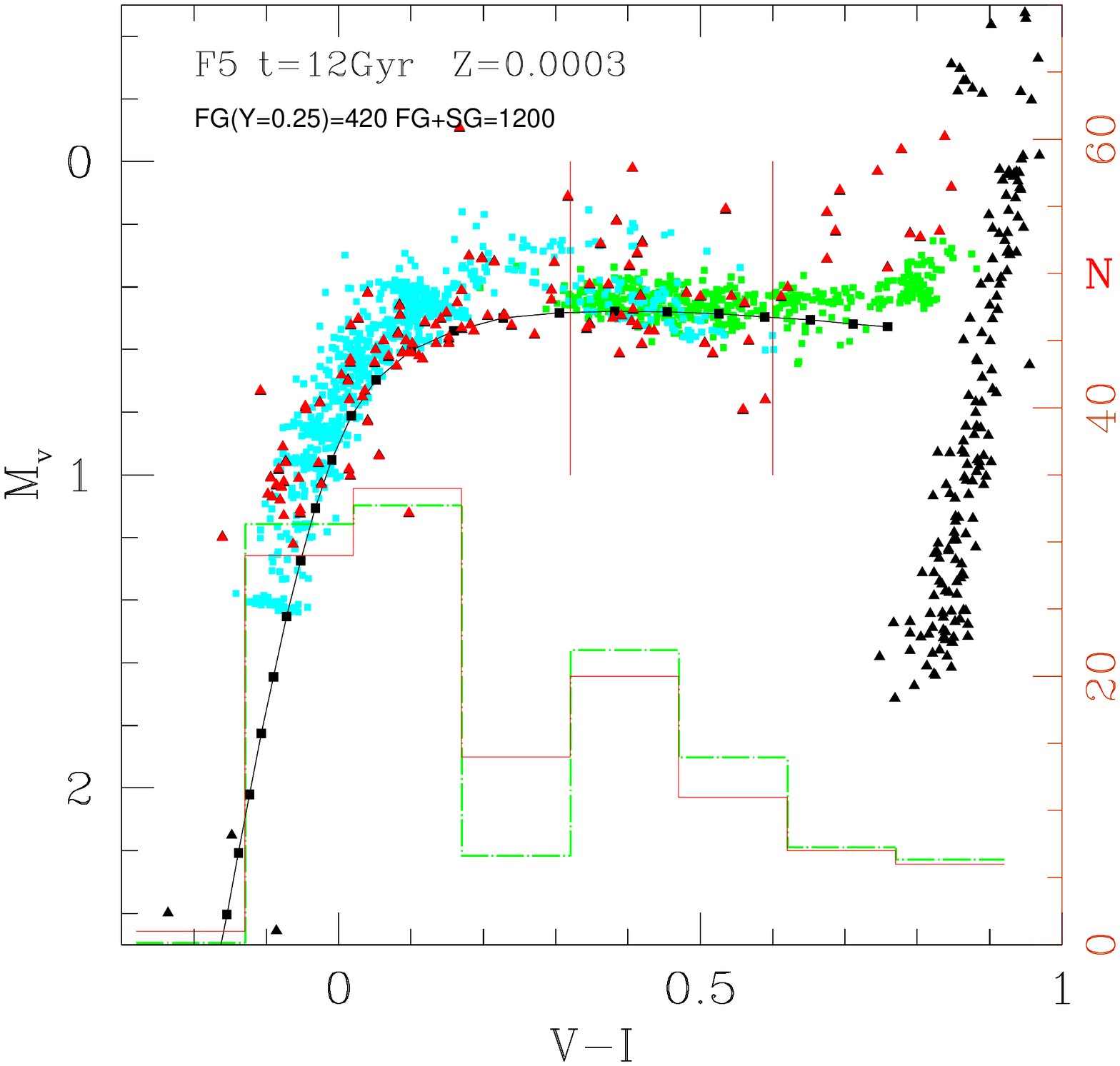}
\includegraphics[width=5cm]{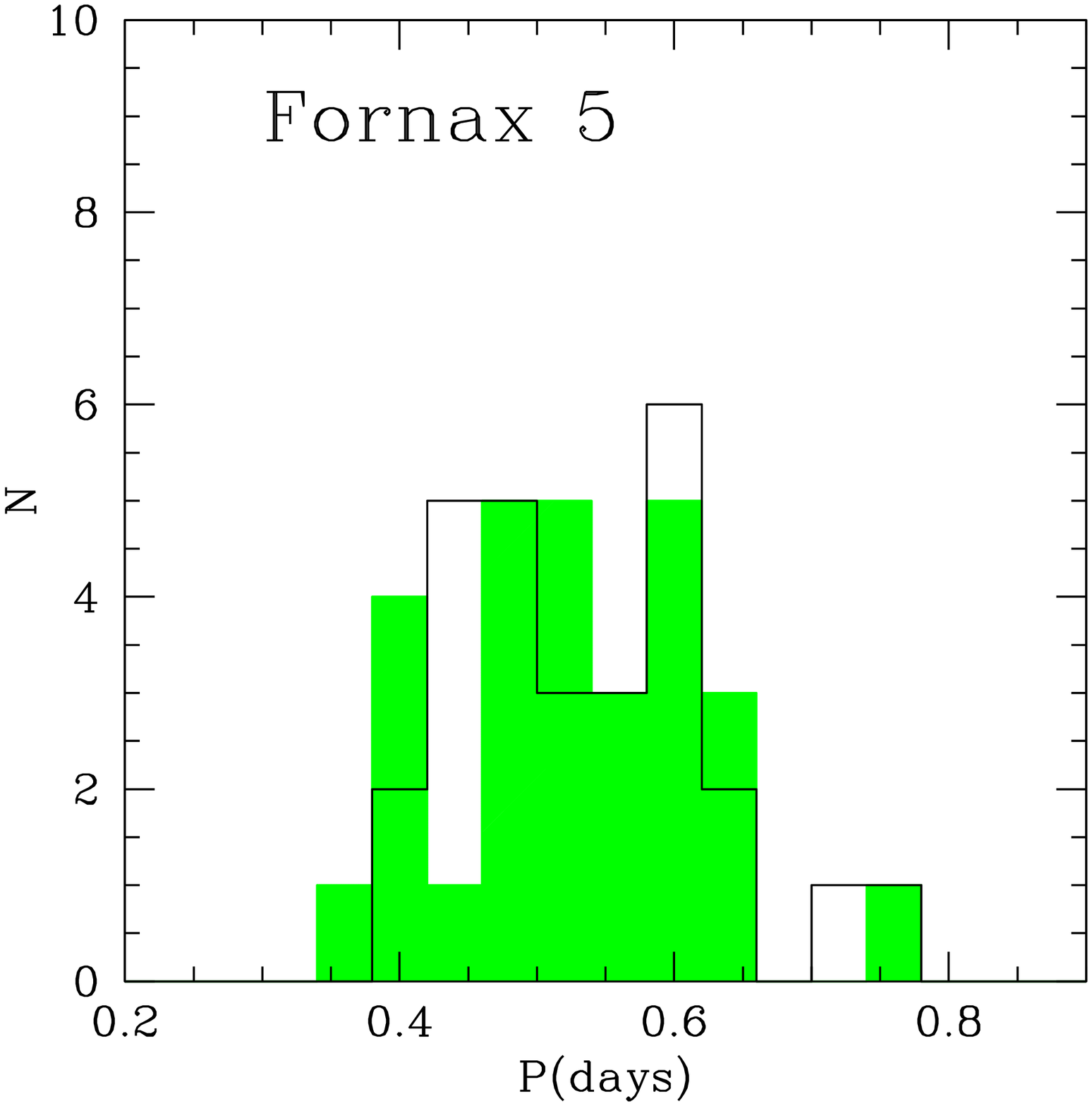}
}
\caption{The left panel shows the result of the simulation model 3 of Table \ref{table2}, where the SG is kept to a minimum; the central panel shows the more satisfactory model 6, and the right panel the observed period distribution from Greco et al. (2009) in the full histogram, compared with the period distribution of a simulation having the inputs of model 6, but for ten times less stars. The left and central panel coding is the same as in Figure \ref{f3}.} 
\label{f4} 
\end{figure*}

Also for F5, listed at [Fe/H]=--2.1 by \cite{larsen2012b}, we use the Z=0.0003 composition. We adopt (V--M$_{\rm v}$)=21.0 and E(V--I)=0.05.  The left panel of Fig.~\ref{f4} shows the result of the simulation model 3, in which we attempt to reproduce the HB by assuming a FG starting its HB
evolution at V--I$\simeq$0; a larger helium content is assumed for a SG
reproducing only the bluer and fainter stars. The FG percentage for this choice
is  60\%. The simulation, however, fails to reproduce the numerous
RR Lyr stars, and the red side of the HB, as in F2. We then resort to a
reduced mass loss on the RGB (0.077\msun), so to reproduce the peak in the RR
Lyrs by the evolution of the FG stars in their longest burning phase, and adopt
larger helium content (SG signature) for all the stars bluer than (V--I)$_0
\sim$0.3 (Fig.\ref{f4}, central panel). The percentage of FG stars is then
reduced from 60\% of the first simulation to only 35\%. The  distribution
obtained for the RR Lyr periods is compared with the data by \cite{greco2009}
in the right panel of Fig.\ref{f4}. This simulation has been performed by
adopting the same number of HB stars as available in the sample, and the total
number of RR Lyrs obtained is $\sim$30, like in the observations. If the
distribution of periods is obtained by adopting ten times more stars (as in the
simulation shown in the central panel), there appears a tight peak in the period distribution, similar to what we
see in the cluster M3, where the number of RR Lyr is very large \citep[about 200, see][]{caloi2008}.
%Notice that the total number of RR Lyr is 30, while the simulation has been made for a much larger number of objects.

\subsection{Simulation of the HB in F3}
\label{F3}
\cite{larsen2012b} assign [Fe/H]=--2.3$\pm$0.1 to this cluster. The
metallicity corresponding to [Fe/H]=--2.3 is  Z=0.95--1.2$\times 10^{-4}$\ for
\alfafe=0.2, or Z=1--1.7$\times 10^{-4}$\ for \alfafe=0.4. The case Z$\sim
10^{-4}$\ is to be excluded for the reasons discussed in Sect.\ref{dating},
since the colour distribution of HB stars has a local maximum in the RR~Lyr
region, like in F2 and F5; a value in the upper range (e.g. Z=1.5$\times
10^{-4}$) would allow a meaningful simulation. Also for this cluster we adopt models with Z=0.0003, as we are mainly interested in the percentage of SG stars, and this is scarcely dependent on the exact metallicity.

\begin{figure}    
\centering{
\includegraphics[width=8cm]{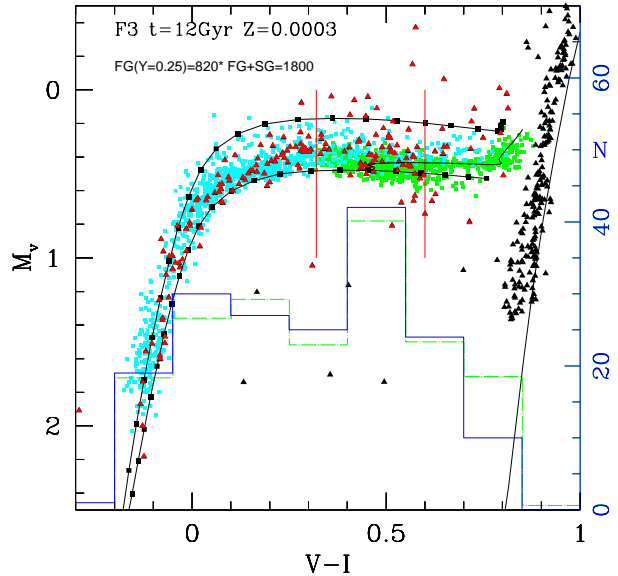}
}
\caption{The result for F3 of simulation 5 in Table \ref{table2}. The red triangles represent the observed data, while the green squares are the FG stars and the cyan squares are the SG stars. The ZAHBs for Z=0.0003 and Y=0.25 (lower line with squares), Y=0.35 (top line) are also shown.} 
\label{f5} 
\end{figure}

Figure \ref{f5} shows one satisfactory simulation (model 5 in Table~\ref{table2}). For an age of 12Gyr, we have set
$\Delta$M=0.0713\msun\ to locate most of the FG in the RR~Lyr region, and we have added several SG samples, with increasing helium up to Y=0.34, in order to account for the blue part of the HB that extends to \teff\ larger than for the other clusters. The percentage of FG stars is here $\sim$46\%.  One may wonder why  the cluster with the largest extension of the HB has in fact a slightly smaller percentage of SG.  The main difference with the cases of F2 and F5 is the following: in F3, although we have also to account for a fraction of stars with more extreme colors,  the most prominent number of stars in the HB is at the RR Lyr colour location, occupied mainly by the FG. Notice that the simulation shows also  SG (evolved) stars in the RR Lyr region: this should be reflected in their period distribution, that is not yet available.  Finally, notice that again we had to add a $\delta \rm{M}_{SG}$=0.03\msun\ to the mass lost in the RGB evolution by SG stars.

\begin{figure*} 
\centering{ 
\includegraphics[width=18cm]{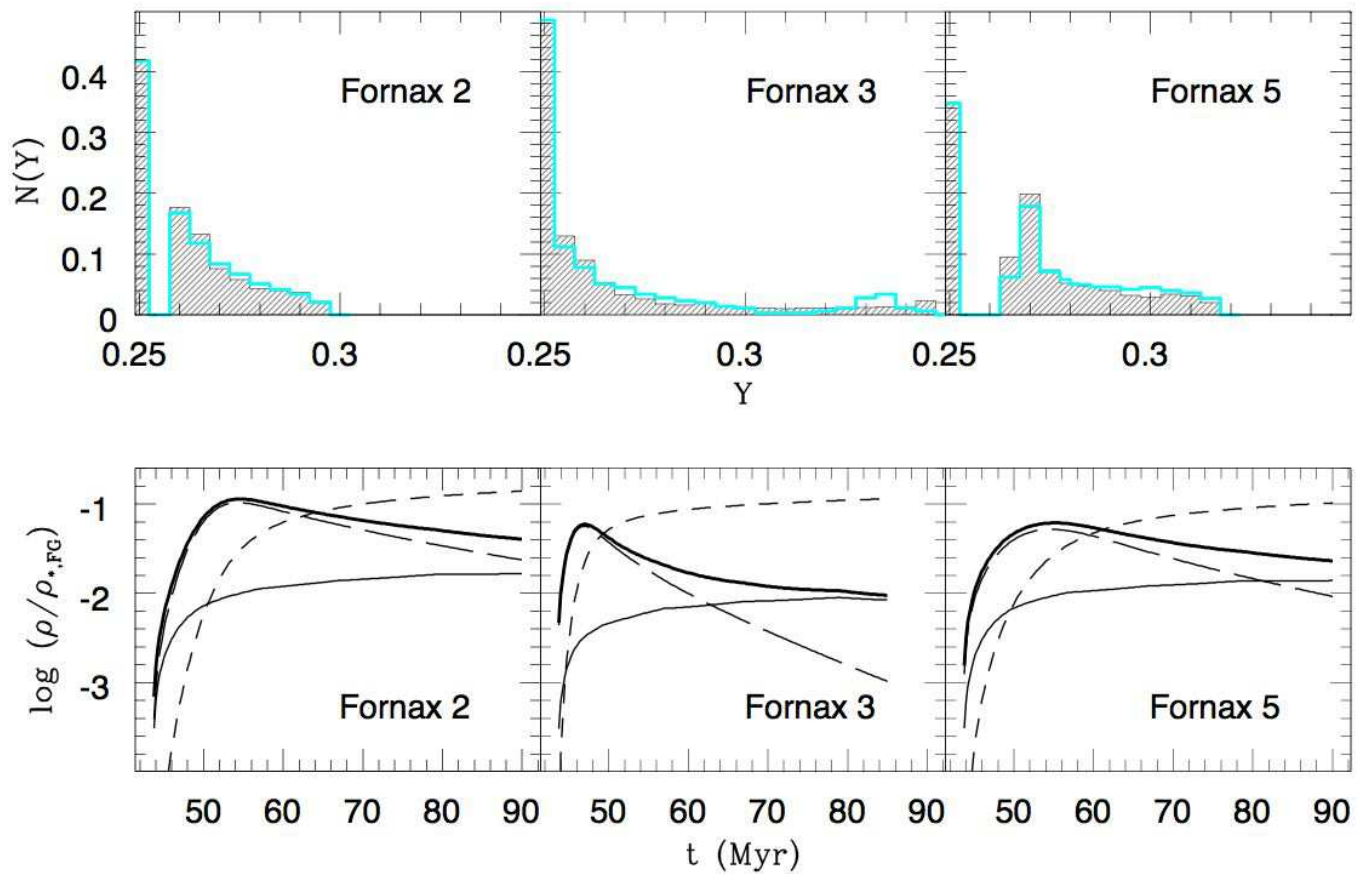}
 } 
\caption{The upper panels show the helium distributions of the three clusters F2, F3 and F5 obtained from the HB morphology (cyan, solid line histograms) and from one-zone chemical models (shaded histograms). The lower panels illustrate the time evolution of different quantities given by the one-zone models: total amount of gas (thick solid line), stellar ejecta (thin solid line), pristine gas (long-dashed line), SG stars (dashed line).}
\label{f6} 
\end{figure*}

\section{Application of the chemical evolution model and the initial
  mass of F2, F3 and F5}
\label{chemevol}
In order to obtain a guess on the initial (pure FG) mass of the
clusters, we resort to our chemical evolution model described in
D2010. This is a one-zone model in which the interstellar gas (ISM) is
supposed to be initially absent owing to the clearing action of the FG
Type II supernovae explosions; the ISM is successively supplied by the
FG AGB ejecta and by the accretion of pristine gas, and is depleted by
the SG star formation. The model is characterized by the initial FG
density $\rho_{\rm *,FG}$ and by a gaussian temporal profile of the
pristine gas accretion described by the density $\rho_{\rm 0,pr}$, the
time $t_{\rm ac}$ at which the maximum accretion occurs, and by the
timescale $\tau$, regulating the with of the gaussian. Another temporal
parameter is given by $t_{\rm end}$, the time at which the simulation
ends. Finally, the star formation (SF) efficiency is
regulated by the parameter $\nu$ (see Eq. 5 in D2010) with values in the range $0<\nu <1$.  
A further parameter is also present: the ratio $x=\rho_{\rm
  *,SG}/\rho_{\rm *,tot}$ between the nowadays alive SG and total
(SG+FG) stars; however, this parameter does  
 not enter directly into the model, but is inferred {\it a posteriori}
 for a more realistic fit of the data (see below).   
 
The models presented here are aimed at reproducing the N(Y)
distributions shown in 
Fig. \ref{f6} (corresponding to the models 4 (F2), 5 (F3) and 6 (F5) of
Table \ref{table2}). Figure \ref{f6} shows the N(Y)
obtained by three chemical models of F2, F3 and F5, characterized by
the following parameters: ($t_{\rm 
  ac,7},t_{\rm end,7},\tau_7,\rho_{\rm FG},\rho_{\rm
  0,pr},\nu,x$)=(5, 9, 0.4, 80, 0.14, 0.08, 0.58), ($t_{\rm ac,7},t_{\rm
end,7},\tau_7,\rho_{\rm FG},\rho_{\rm
0,pr},\nu,x$)=(4.5, 8.5, 0.2, 300, 0.1, 0.2, 0.52), and ($t_{\rm ac,7},t_{\rm
end,7},\tau_7,\rho_{\rm FG},\rho_{\rm
0,pr},\nu,x$)=(48.5, 9, 0.7, 85, 0.1, 0.13, 0.65), respectively; here the
times are expressed in $10^7$ yr, $\rho_{\rm FG}$ in M$_{\odot}$
pc$^{-3}$ and $\rho_{\rm 0,pr}$ is normalized to $\rho_{\rm FG}$. We
stress that the adopted values of $x$ are not arbitrary, but are
obtained from the models 4, 5 and 6 (see Table \ref{table2}). The
lower panels in Figure \ref{f6} also show the time evolution of the
amount of AGB ejecta, pristine gas and SG stellar mass given by the
one-zone models. The SG formation starts at an age of 44~Myr, at which the Type II supernova epoch ends in the adopted models by \cite{ventura2013} of Z=0.0003,  and the cooling flow begins to collect the super--AGB ejecta in the cluster core. 

The total initial mass of SG stars formed and the ratio of this mass to the initial
FG mass in the models that can reproduce the N(Y) in the three
clusters can be estimated by looking at the final values of the dashed
curves in the lower panels of Fig. \ref{f6}. 
These values along with the current observed fraction of SG
stars allow to estimate the ratio the total initial to current GC
mass. The estimate will also depend, of course, on the fraction 
of the initial SG stars still alive.

From Table 2 we have the fraction of SG stars derived from the HB simulations. Although the observed ratio can be dependent on the detail of the dynamical mixing suffered by the two populations, and thus on the location of photometric sample in the clusters \citep{vesperini2013}, we take it as a proxy of the fraction of SG mass alive today, with respect to the total GC mass alive today.

\begin{equation}
{{\rm N}_{\rm SG} \over {\rm N}_{\rm tot} }=  { M_{\rm SG~alive,~now} \over  M_{\rm GC~alive,~now} }
\end{equation}

We choose 0.1\msun\ as minimum mass entering in the initial mass function, and consider as ``alive" the stars between 0.1 and 0.8\msun, both for the FG and SG. Throughout this discussion, we apply the stellar mass function by Kroupa 2001.

The cumulative mass of the SG in units of the initial (FG only) mass is shown in Fig. \ref{f6}. Its final value in the models reproducing the helium distribution of the three cluster is $\sim$0.1 in the three successful simulations. Therefore we have:
\begin{equation}
{\rm M}_{\rm FG} \simeq 10 {\rm M}_{\rm SG}.
\label{eqmfg}
\end{equation}
\noindent
While for ${\rm M}_{\rm FG}$\ entering the chemical evolution models we made the hypothesis that it was initially distributed in stars from 0.1 to 100\msun, we still have to consider the mass function of the SG stars. 
%REFEREE
The possible formation of stars in a cooling flow, and the shape of their IMF, is a long-standing problem linked to the more general conundrum of the final physical disposition of the cooled gas. Formation of only low mass stars has been suggested in order to explain the absence of the huge amounts of cold gas expected at the centre of the cooling flows. Currently, there is no conclusive evidence either for or against such an hypothesis (see, e.g. Kroupa \& Gilmore 1994, Mathews \& Brighenti 1999 and references therein).
%REFEREE
In the context of the formation of SG stars in GCs, some works \citep{prantzos2006,dercole2008} have considered that only low mass stars are formed from the SG gas, from 0.1 to 0.8\msun. This assumption minimizes the requirements on the initial FG mass, as all the SG mass would still be in stars alive today. A constraint on the SG mass function comes from the assumed AGB scenario: the forming SG stars should not more massive than the minimum mass for supernova explosion, say 8 \msun, otherwise the stellar explosions would blow out the AGB ejecta, preventing its accumulation within the cluster and any further star formation (e.g. D2012). In this case, only 50\% of the mass is today contained in alive stars between 0.1 and 0.8\msun.

So we have ${\rm M}_{\rm SG}= q \times {\rm M} _{\rm SG~alive,~now}$, with {\it q} in the range 1--2.

We are interested in the ratio of the initial GC mass (M$_{\rm FG}$ in equation \ref{eqmfg}) with respect to the present cluster mass M$_{\rm GC,~now}$. We can write then

\begin{equation}
{\rm M_{\rm FG} \over {\rm M}_{\rm GC,~now} }  \simeq 10 \times q \times { \rm M_{\rm SG~alive,~now} \over  \rm M_{\rm GC~alive,~now} }\times {{ \rm M_{\rm GC ~alive,~now} \over  \rm M_{\rm GC,~now}}}, 
\label{finalperc}
\end{equation}

From \cite{kroupa2001}, the alive stars constitute a fraction $\approx$ 0.7 of all the initial stars; this fraction can actually be lower since dynamical evolution and mass loss due to two-body relaxation may reduce this ratio due to the preferential loss of low-mass main sequence stars (see e.g. Vesperini \& Heggie 1997, Giersz 2001, Baumgardt \& Makino 2003). Substituting in equation \ref{finalperc} the values of ${ \rm M_{\rm SG alive,now} \over  \rm M_{\rm GC alive,now} }$\ from Table 2, we get that the initial mass of the clusters were about $q \times 3.5-5$ times larger than their current mass. If only half of the initial SG stars are still alive ({\it q}=2) the initial mass of the clusters must have been about 7--10 times larger than the current mass. We emphasize that different models (see the example Fornax 2a in the
Appendix) indicate that such a factor is not affected by the value of the SF efficiency $\nu$; in fact, this value is regulated by the need to reproduce the helium distribution, but it is not important in the determination of the final mass of the SG population whenever the replenishing timescale of the accreting gas (AGB ejecta and pristine gas) is shorter than the cluster evolution time  \citep{dercole2008}, as indeed is the case in the present models. 
On the other hand, in the absence of further observational constraint (such as, for example, a well definite O-Na anticorrelation), different set of parameters in the chemical evolution models can give similar, reasonable fit of the helium distributions. In particular, models exist with a shorter evolutive time which give rise to N(Y) close to those shown in Fig. \ref{f6}; although rather extreme, these models keep open the option of a ratio
${ \rm M_{\rm FG initial} \over  \rm M_{\rm GC now} }$  as high as $\sim 20-25$ (see the Appendix).
Reversely, we can not increase the mass of the SG by extending the star formation beyond 90--100 Myr, as, at later times, we expect that the onset of type I supernova explosions stops the cooling flow, and, further, the composition of AGB ejecta becomes incompatible with the chemical patterns seen in the GCs of our Galaxy \citep{dercole2008}. We conclude that the initial mass of the clusters F2, F3 and F5 in Fornax is contained within a factor 7--10 of the present masses.

This value is not very different from what we envisaged in \cite{dercole2008}, and more explicitly modeled in \cite{vesperini2010}. 

A combination of different model ingredients (much narrower range of FG star  masses contributing to the gas for SG formation, a shorter duration of the SG formation episode, use of only a small fraction of the polluted gas, lack of pristine gas) may result in the much larger mass factors cited in the literature and discussed in the Introduction.
\section{Discussion}

There are some critical points in the interpretation of these extra--Galactic GCs that deserve discussion. 

This investigation was prompted by the constraint on the possible initial mass of GCs set  by \cite{larsen2012a} on the basis of the observed number of very metal poor field stars in Fornax.
As pointed out by \cite{larsen2012a}, the currently known number of
very metal poor field stars in Fornax appears to limit the initial
mass of GCs in Fornax to be at most 5-6 times its current
mass. Interestingly, their analysis also suggests  that the initial
Fornax cluster system could not include any additional clusters that have now
completely dissolved.

Our simulations of the HBs of Fornax clusters suggest that one of them
(F1) hosts only the FG component; the lack of SG stars implies that no
constraints on its initial mass can be set on the basis of the SG
formation and dynamical evolution history. The cluster F4 instead is assigned by
recent determinations to a metallicity range ([Fe/H]$\sim$--1.4) where stars
in Fornax field are in large number, and no strong limit on F4 initial
mass  is imposed by field stars.

However, the other three clusters (F2, F3, F5: all very metal poor)
harbour a substantial SG population. Our models show that an initial
mass  4--5 times the present mass is possible, if we impose that all SG stars are alive
today. By adopting an initial mass function of the SG stars extended
up to $\sim$8 \msun, this estimate must be raised by a factor
$\sim$2. A further factor 2.5 is necessary if the time of formation of
the SG is shorter than that of the one-zone models shown in
Fig. \ref{f6}. We consider a factor of about 25 as an upper limit.

Although our  lower limits are consistent with the constraint set by the study of \cite{larsen2012a}, it is important to understand how robust such a constraint is.

One possible major issue in using the field stars to constrain the birth mass of GCs resides in the estimate of the metallicity of field stars and of GCs.

In this regard, we stress again the impossibility of fitting an isochrone of the appropriate age to a Fornax cluster with [Fe/H]  $\sim$ --2.5 dex: this estimate should be raised at least to --2.3 or --2.2, depending on the solar metallicity. On the contrary, the CM diagram morphology of cluster F4 does not appear compatible with the factor 10 increase in [Fe/H] with respect to F1, reported by recent spectroscopic determinations (see Fig.~2). 

 As for field metallicity, it is based on Ca--triplet calibration which implies errors of at least 0.1 -- 0.2 dex
\citep{battaglia2008, starkenburg2010}; besides, the data for Fornax field contain only three stars with [Fe/H] $<$ --1.00, of which {\it   one} has [Fe/H]$<$ --2.00, making a reliable calibration difficult. This lack of data for the very metal poor component may arise from a real scarcity, or, as suggested by Starkenburg et al. (2010), from the very small fraction of field RGs sampled at present. They remark also that the very low metallicity tail of Fornax is different both in shape and number from the other classical dSphs and the halo, and comment that the dominant young, metal rich population may well be hiding the scarce metal poor component.     

In closing we point out that the use of field stars to constrain the initial GC masses is strongly undermined if Fornax has lost a fraction of its field stars and/or if its globular clusters formed in a different system and were were later accreted in Fornax (see Larsen et al. 2012 and references therein).

\subsection{A difference in mass loss on the RGB for the FG and SG components? }
\label{F6}
A difficulty was met when considering the blue side of the HBs in F2, F3 and F5. As an example, we shall consider F3: its ``blue'' population can not be imputed only to an increased helium. The location of the ``zero age" HB would follow the
green line in Fig. 7, if we do not allow for an extra mass loss during the RGB evolution of the SG population.
A simulation taking into account only the effect of helium variation in
decreasing the evolving mass would follow the green line, and would predict a
distribution of stars along and above it, while the region between it and the
Y=0.25 ZAHB is well populated. The blue line shows the ZAHB obtained taking into account an extra--mass loss of 0.03\msun: this is enough to shift the ZAHB to a location much closer to the lower envelope of the data.

A similar situation is found in very metal poor Galactic GCs with a similar distribution of HB stars, including a well populated red, variable and blue regions. As mentioned in \cite{dantonacaloi2008}, the RR~Lyraes and the red HB of M~68 can be nicely fit by a FG with a small mass spread, as found in the case of M~3 \citep{caloi2005}. On the contrary, the simulated luminosity of the blue side results too large if interpreted only in terms of an increased helium content. The same considerations apply to the HB in M15, another very metal poor GC rich in blue, variable and red HB stars.

This difficulty is less evident in the case of the more metal rich (Z$\sim 10^{-3}$) Galactic GCs analysed up to now, but hints that a larger mass loss is required for the SG come from NGC~1851 \citep{salaris2008} and from the analysis by \cite{dalessandro2013} of the HBs in M~3, M~13 and M~79.

\begin{figure}    
\centering{
\includegraphics[width=8cm]{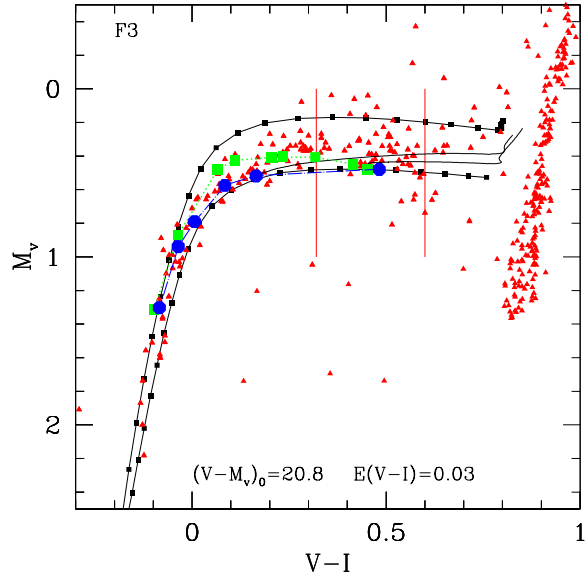}
}
\caption{Data for F3 are shown as red triangles. Green squares indicate the 
ZAHB with enhanced Y assuming the same mass loss as for Y=0.25. An extra mass
loss of 0.03\msun\ gives rise to the ZAHB indicated by blue circles, much
closer to the observations. The ZAHBs for Y=0.25 and Y=0.35 are also shown.} 
\label{f7} 
\end{figure}

An increase in mass loss from SG giants may be related to an average larger angular velocity in SG main sequence stars with respect to FG ones. In current views, SG population takes origin from a cooling flow collecting at the cluster centre and so the initial number density of SG MS stars is larger than in the FG case. We may speculate that such an increased density favours  an increase in the average rotational velocity of the newly formed stars. The effects of an initial, relatively slow angular rotation in Pop II main sequence stars ($\omega_{MS}$) has been studied, in first approximation, in the years '70s by \cite{mengel-gross1976} and \cite{renzini1977}. The result was that, as long as $\omega_{MS}$ does not exceed 2 $10^{-4}$s$^{-1}$, the turnoff age is not altered, while both the luminosity at the He--flash and the He--core mass increase  (eqs. 2.19 and 2.20 in Renzini's paper). On the basis of \cite{fpr1975} treatment of mass loss along the RG branch, Renzini gives 
 an estimate of the relation between the increase in luminosity at the RG tip and the related increase in mass loss. A mass loss increase of 0.03 \msun$~$requires an increase in ${\log(L/L_\odot)}$ at the RG tip of 0.01, in turn requiring an $\omega_{MS}$\ of about 1.2 $10^{-4}$s$^{-1}$. The expected increase in the core mass is 0.004 \msun. Such an increase does not affect substantially the ZAHB location, whose luminosity  would be {\it larger} by only $\sim$0.03~mag.\footnote{We use the relation ${\delta \log {\rm L}/L_\odot} \over {\delta {\rm M}_{core}/M_\odot}$=3 from \cite{sweigart-gross1976}.} The constancy of the assumed mass loss increase for the SG population implies a small dispersion in $\omega_{MS}$, an approximation in line with the qualitative treatment exposed here. We stress that the hypothesis of an increased  average rotation is just a suggestion, in view of future investigations on the subject.

A final remark on the stellar components of Fornax field. According to our
interpretation of the HBs, at least three GCs (F2, F3 and F5) host  a large
population of SG stars, mostly located on the blue side of the HB. From the
exemplifications of Figs.~\ref{f3} (left panel), \ref{f4} (central panel) and
\ref{f5}, we qualitatively see that the FG stars do not extend much to the
blue side of the RR Lyr gap (in the cases displayed, there are no FG stars in
the HB at (V--I)$_0 \simlt$0.3). If our interpretation of the HB morphology is
correct, it seems therefore that these objects can form {\it only}
in the enviroment of a GC, from which they populate the galactic field when
they are lost by the native cluster. As discussed in \cite{vesperini2013}, SG star evaporation from GCs into the host galaxy field would occur during a cluster long-term evolution driven by two-body relaxation. 

We analyzed the data in de Boer et al. (2012), who
%(2012, Fig. 9, last panel), we  derive a ratio between SG (BHB stars) and FG (RR Lyr variables and RHB stars) of 
present deep optical photometry of the Fornax dSph in the B, V and I filters
obtained using the CTIO 4-m MOSAIC II camera. They define an elliptical radius
\rell\ to provide a distance measurement from the center. Since the spatial
coverage of the B and V filters is complete for \rell$\simlt$0.8 degrees,
while the I filters is complete for \rell$\simlt$0.4 degrees we have used B
and V bands data with 0.5$<$\rell$<$0.8 in order to be able to isolate mostly
the old population. In fact de Boer et al. 2012 have shown that the young and
intermediate age stars are more centrally concentrated than the old
stars. 
%NEWMOD
The data we use do not give information on stellar metallicity, but a combination of the spectroscopic catalogs of DART (CaT and HR), the medium resolution spectra catalog by \cite{kirby2010} (med. res) and the CaT data by \cite{pont2004}, kindly done by K. de Boer (private communication) shows that about half of the stars at 0.5$<$\rell$<$0.8 have metallicity [Fe/H]$<$--1.4, while most of the younger stars of [Fe/H]$>$--1.0 are confined within \rell$<$0.4). 
%NEWMOD
We find that this region of Fornax field shows a well
populated sample of stars  hotter than the  blue edge of the RR Lyrae strip  (fixed at (B--V)$_0$=0.2mag) and make up 9--11\% of the whole  HB population (assuming E(B-V)=0.03mag). 
%NEWMOD
If we assume that all the stars having  [Fe/H]$>$--1.4 populate the red HB only, the percentage of blue HB stars in the low metallicity sample is $\sim$20\%. 

A further detailed analysis of the HB field population would be very important to shed light on the possible connection between multiple stellar populations in globular clusters and in the Fornax field stars.

\section*{Acknowledgments}
We thank Thomas de Boer and Eline Tolstoy for sharing their photometry of Fornax field stars. We also thank Gisella Clementini for useful discussion on the RR Lyrs in the Fornax clusters, and Carlo Corsi for making available the clusters photometric data.

F.D., A.D and P.V. have been supported by the PRIN--INAF 2011 ``Multiple populations in globular clusters: their role in the Galaxy assembly", P.I. E. Carretta. EV was supported in part by grant NASA-NNX13AF45G.

\appendix
\section{Chemical model analysis }
\label{sec:app}

\begin{figure*} 
\centering{ 
\includegraphics[width=18cm]{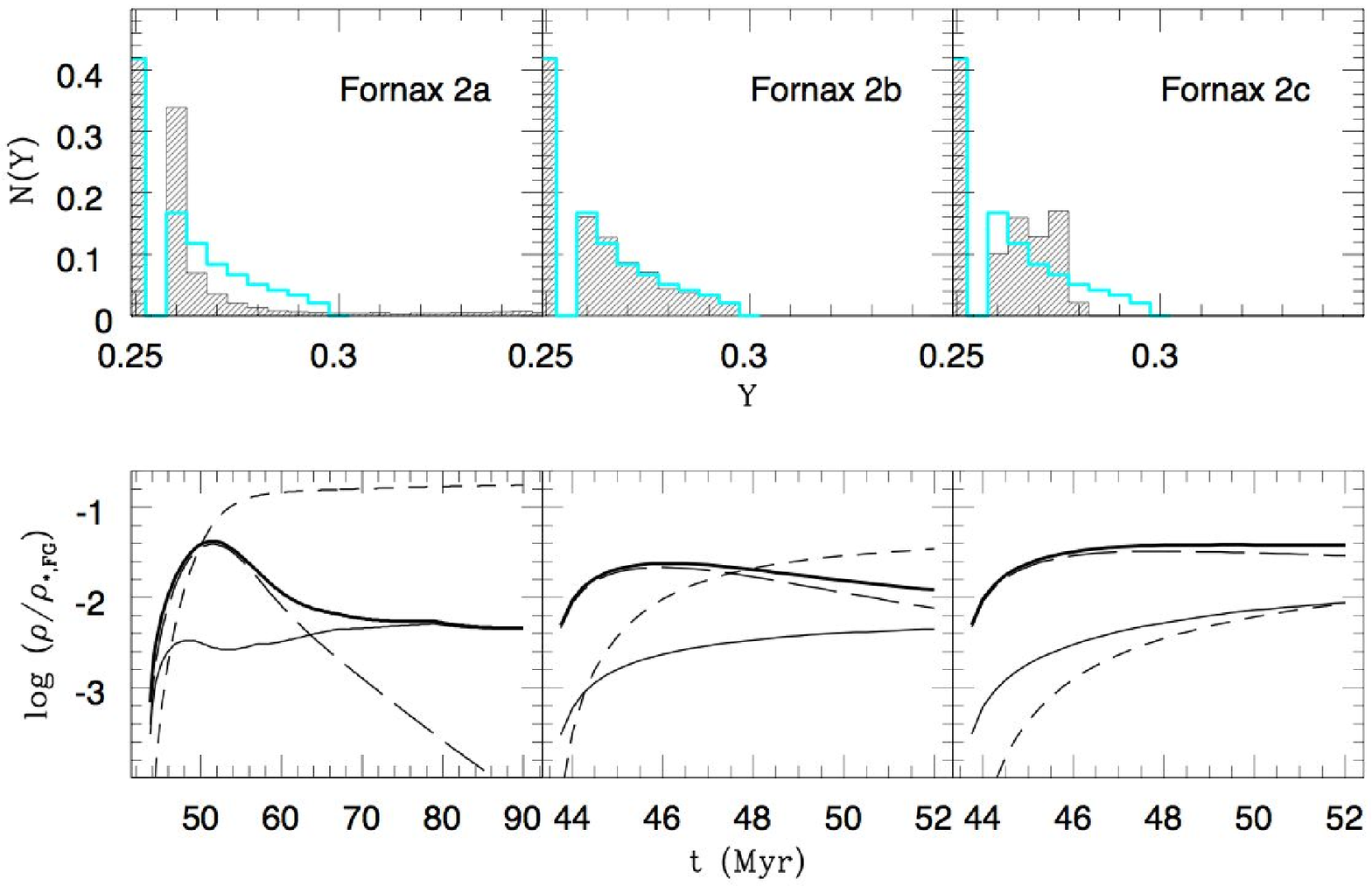}
 } 
\caption{The symbols are the same as for for Fig. \ref{f6}. The figure shows three cases of chemical evolution for F2. The left panel shows the difference with the standard case, when we assume star formation efficiency $\nu$=1: the helium distribution is different, but the total mass in SG is barely larger than in the standard case having $\nu$=1 because the accretion time is shorter than the evolutionary time. The central panel shows a good reproduction of N(Y) achieved by a model in which the SF formation has a short duration (up to 52 Myr) and $\nu$=1. The right panel shows the result for the same inputs of the central panel, but $\nu$=0.1. In this case, the total amount of gas in SG stars is a factor 2--3 smaller, as the accretion time is now comparable to the evolutionary time. }
\label{figc2diff} 
\end{figure*}

In this Appendix we discuss in a little more in detail the response of our one-zone model to the variation of the SF efficiency $\nu$ and to the duration of the evolutive time.
In all the models presented in the text (Fig. \ref{f6}), the SG formation lasts until a total age of 85--90 Myr. As the only constraint is the helium distribution function, this choice is not unique. As an example, Figure \ref{figc2diff} shows three different cases for the cluster F2: in the first case (left panel, Fornax 2a) we show that the total amount of gas in SG stars barely increases when we shift from $\nu$=0.08 (standard model for F2, in Fig. \ref{f6}) to $\nu$=1, because, as discussed by \cite{dercole2008}, the accretion time is shorter than the evolutive time. Other models are possible to reproduce N(Y), and the central panel shows an extreme model, with $\nu$=1, in which the star formation lasts for only 8 Myr (Fornax 2b). Obviously, the mass in SG stars is now a much smaller fraction of the FG mass; we have $\rho_{\rm SG} \simeq 0.04 \rho_{\rm FG}$, leading to a ratio FG$_{\rm initial}$/(FG+SG)$_{\rm now} \sim 30$. If we further reduce the star formation efficien
 cy to $\nu$=0.1 (Fornax 2c), the total mass achieved in SG stars is even smaller, because now the accretion time is comparable to the evolution time, and not all the gas collected in the cluster core can be used to form stars in such a short time. 

We regard the case Fornax 2b as an extreme example still able to obtain a satisfactory N(Y), but only other information on the chemistry of SG stars can improve the fit with observations and allow us to choose among different models. For now, we use the models of Fig. \ref{f6} as standard models for the three clusters.

\label{lastpage}

\begin{thebibliography}{}
\bibitem[Angulo et al. (1999)]{angulo} Angulo C., Arnould M., Rayet M., et al. 
1999, Nucl. Phys. A, 656, 3 
\bibitem[Asplund et al.(2009)]{asplund2009araa} Asplund, M., Grevesse, N., Sauval, A.~J., \& Scott, P.\ 2009, \araa, 47, 481 
\bibitem[Bahcall et al.(2001)]{bahcall2001} Bahcall, J.~N., Pinsonneault, M.~H., \& Basu, S.\ 2001, \apj, 555, 990 
\bibitem[Bahcall et al.(2004)]{bahcall2004} Bahcall, J.~N.,  Serenelli, A.~M., \& Pinsonneault, M.\ 2004, \apj, 614, 464 
\bibitem[Bahcall et al.(2005)]{bahcall2005} Bahcall, J.~N., Basu, S., \& Serenelli, A.~M.\ 2005, \apj, 631, 1281 
\bibitem[Bahcall et al.(2006)]{bahcall2006} Bahcall, J.~N., Serenelli, A.~M., \& Basu, S.\ 2006, \apjs, 165, 400 
\bibitem[Basu \& Antia(2004)]{basu2004} Basu, S., \& Antia, H.~M.\ 2004, \apjl, 606, L85 
\bibitem[Battaglia et al.(2006)]{battaglia2006} Battaglia, G., Tolstoy, E., Helmi, A., et al.\ 2006, \aap, 459, 423 
\bibitem[Battaglia et al.(2008)]{battaglia2008} Battaglia, G., Irwin, M., Tolstoy, E., et al.\ 2008, \mnras, 383, 183 
\bibitem[Baumgardt \& Makino(2003)]{baumgardt2003} Baumgardt, H., Makino, J., 2003, \mnras, 340, 227
\bibitem[Bekki et al.(2007)]{bekki2007} Bekki, K., Campbell, S.~W., Lattanzio, J.~C., \& Norris, J.~E.\ 2007, \mnras, 377, 335 
\bibitem[Bekki(2011)]{bekki2011} Bekki, K.\ 2011, \mnras, 412, 2241
\bibitem[Buonanno et al.(1998)]{buonanno1998} Buonanno, R., Corsi, C.~E., Zinn, R., et al.\ 1998, \apjl, 501, L33 
\bibitem[Buonanno et al.(1999)]{buonanno1999} Buonanno, R., Corsi, C.~E., Castellani, M., et al.\ 1999, \aj, 118, 1671 
\bibitem[Caloi \& D'Antona(2005)]{caloi2005} Caloi, V., \& D'Antona, F.\ 2005, \aap, 435, 987 
\bibitem[Caloi \& D'Antona(2008)]{caloi2008} Caloi, V., \& D'Antona, F.\ 2008, \apj, 673, 847 
\bibitem[Caloi \& D'Antona(2011)]{caloi2011} Caloi, V., \& D'Antona, F.\ 2011, \mnras, 417, 228 
\bibitem[Carretta \& Gratton(1997)]{cg1997} Carretta, E., \& Gratton, R.~G.\ 1997, \aaps, 121, 95 
\bibitem[Carretta et al.(2009a)]{carretta2009a} Carretta, E., et al.\ 2009a, \aap, 505, 117 
\bibitem[Carretta et al.(2009b)]{carretta2009iron} Carretta, E., Bragaglia, A., Gratton, R., D'Orazi, V., \& Lucatello, S.\ 2009b, \aap, 508, 695 
\bibitem[Cassisi et al.(2007)]{cassisi2007} Cassisi, S., Potekhin, A.~Y., Pietrinferni, A., Catelan, M., \& Salaris, M.\ 2007, \apj, 661, 1094 
\bibitem[Christensen-Dalsgaard et al.(1996)]{cd1996} Christensen-Dalsgaard, J., Dappen, W., Ajukov, S.~V., et al.\ 1996, Science, 272, 1286 
\bibitem[Cole et al.(2004)]{cole2004} Cole, A.~A., Smecker-Hane, T.~A., Tolstoy, E., Bosler, T.~L., \& Gallagher, J.~S.\ 2004, \mnras, 347, 367 
\bibitem[Da Costa \& Armandroff(1995)]{dca1995} Da Costa, G.~S., \& Armandroff, T.~E.\ 1995, \aj, 109, 2533 
\bibitem[Da Costa \& Hatzidimitriou(1998)]{dch1998} Da Costa, G.~S., \& Hatzidimitriou, D.\ 1998, \aj, 115, 1934 
\bibitem[Dalessandro et al.(2013)]{dalessandro2013} Dalessandro, E., Salaris, M., Ferraro, F.~R., Mucciarelli, A., 
\& Cassisi, S.\ 2013, \mnras, 430, 459 
\bibitem[D'Antona et al.(2002)]{dantona2002} D'Antona,F., Caloi, V., Montalb{\' a}n, J., Ventura, P., \& Gratton, R.\ 2002, A\&A
395, 69  
\bibitem[D'Antona \& Caloi(2008)]{dantonacaloi2008} D'Antona, F., \& Caloi, V.\ 2008, 
MNRAS, 390, 693
\bibitem[de Boer et al.(2012)]{deboer2012} de Boer, T.~J.~L., Tolstoy, E., Hill, V., et al.\ 2012, \aap, 544, A73 
\bibitem[Decressin et al.(2007)]{decressin2007} Decressin, T., Charbonnel, C., \& Meynet, G.\ 2007, \aap, 475, 859 
\bibitem[D'Ercole et al.(2008)]{dercole2008} D'Ercole, A., Vesperini, E., D'Antona, F., McMillan, S. L. W., Recchi, S.,2008, \mnras, 391, 825
\bibitem[D'Ercole et al.(2010)]{dercole2010} D'Ercole, A., D'Antona, F., Ventura, P., Vesperini, E., 
\& McMillan, S.~L.~W.\ 2010, \mnras, 407, 854 (D2010) 
\bibitem[D'Ercole et al.(2012)]{dercole2012} D'Ercole, A., et al.\ 2012, \mnras, 423, 1521 (D2012)
\bibitem[Di Criscienzo et al.(2004)]{dicriscienzo2004} Di Criscienzo, M., Marconi, M., \& Caputo, F.\ 2004, \apj, 612, 1092 
\bibitem[Di Criscienzo et al.(2011)]{dicriscienzo2011} Di Criscienzo, M., et al.\ 2011, \mnras, 414, 3381
\bibitem[Dotter et al.(2007)]{dotter2007} Dotter, A., Chaboyer, B., Jevremovi{\'c}, D., et al.\ 2007, \aj, 134, 376 
\bibitem[Ferguson et al.(2005)]{ferguson2005} Ferguson  J. W., Alexander  D. R., Allard  F.  et al., 2005, ApJ, 623, 585
\bibitem[Freeman \& Bland-Hawthorn(2002)]{freeman2002} Freeman, K., \& Bland-Hawthorn, J.\ 2002, \araa, 40, 487 
\bibitem[Fusi-Pecci \& Renzini(1975)]{fpr1975} Fusi-Pecci, F., \& Renzini, A.\ 1975, \aap, 39, 413 
\bibitem[Giersz(2001)]{giersz2001} Giersz, M. 2001, \mnras, 324, 218
\bibitem[Gratton et al.(2012)]{gratton2012rev} Gratton, R.~G., Carretta, E., \& Bragaglia, A.\ 2012, A\&A Rev, 20, 50 
\bibitem[Greco (2007)]{greco2007} Greco, C., 2007, PhD Thesis, XIX Ciclo, Universita' di Bologna
\bibitem[Greco et al.(2009)]{greco2009} Greco, C., Clementini, G., Catelan, M., et al.\ 2009, \apj, 701, 1323 
\bibitem[\protect\citeauthoryear{Grevesse \& Sauval}{1998}]{gs98} Grevesse N., Sauval A.J, 1998, SSrv, 85, 161
\bibitem[Harris \& van den Bergh(1981)]{hvdb1981} Harris, W.~E., \& van den Bergh, S.\ 1981, \aj, 86, 1627 
\bibitem[Hodge(1961)]{hodge1961} Hodge, P.V. 1961, \aj, 66, 83
\bibitem[Iglesias \& Rogers (1996)]{iglesias1996} Iglesias, C.A., Rogers, F.J., 
   ApJ, 464, 943
 \bibitem[Kirby et al.(2010)]{kirby2010} Kirby, E.~N., Guhathakurta, P., Simon, J.~D., et al.\ 2010, \apjs, 191, 352 
 \bibitem[Kroupa et al.(1993)]{kroupa1993} Kroupa, P., Tout, C.~A., \& Gilmore, G.\ 1993, \mnras, 262, 545 
\bibitem[Kroupa(2001)]{kroupa2001} Kroupa, P.\ 2001, \mnras, 322, 231 
\bibitem[Kroupa \& Gilmore(1994)]{kroupagilmore1994} Kroupa P., \& Gilmore G.\ 1994, \mnras, 269, 655
\bibitem[Larsen et al.(2012a)]{larsen2012a} Larsen, S.~S., Strader, J., \& Brodie, J.~P.\ 2012a, \aap, 544, L14 
\bibitem[Larsen et al.(2012b)]{larsen2012b} Larsen, S.~S., Brodie, J.~P., \& Strader, J.\ 2012b, \aap, 546, A53 
\bibitem[Letarte et al.(2006)]{letarte2006} Letarte, B., Hill, V., Jablonka, P., et al.\ 2006, \aap, 453, 547 
\bibitem[Mackey \& Gilmore(2003)]{mackey2003} Mackey, A.~D., \& Gilmore, G.~F.\ 2003, \mnras, 345, 747 
\bibitem[Mathews \& Brighenti(1999)]{mb1999} Mathews W. G., \& Brighenti F.\ 1999, \apj, 527, L31
\bibitem[Mengel \& Gross(1976)]{mengel-gross1976} Mengel, J.~G., \& Gross, P.~G.\ 1976, \apss, 41, 407 
\bibitem[Montalb{\'a}n et al.(2004)]{montalban2004} Montalb{\'a}n, J., Miglio, A., Noels, A., Grevesse, N., \& di Mauro, M.~P.\ 2004, SOHO 14 Helio- and Asteroseismology: Towards a Golden Future, 559, 574 
\bibitem[Montalban et al.(2006)]{montalban2006} Montalban, J., Miglio, A., Theado, S., Noels, A., 
\& Grevesse, N.\ 2006, Communications in Asteroseismology, 147, 80 
%\bibitem[Oh et al.(2000)]{oh2000} Oh, K.~S., Lin, D.~N.~C., \& Richer, H.~B.\ 2000, \apj, 531, 727 
\bibitem[Pont et al.(2004)]{pont2004} Pont, F., Zinn, R., Gallart, C., Hardy, E., \& Winnick, R.\ 2004, \aj, 127, 840 
\bibitem[Potekhin et al.(1999)]{potekhin1999} Potekhin, A.~Y., Baiko, D.~A., Haensel, P., \& Yakovlev, D.~G.\ 1999, \aap, 346, 345 
\bibitem[Prantzos \& Charbonnel(2006)]{prantzos2006} Prantzos, N., \& Charbonnel, C.\ 2006, \aap, 458, 135 
\bibitem[Reimers(1975)]{reimers1975} Reimers, D.\ 1975, Problems in stellar atmospheres and envelopes., (A75-42151 21-90) New York, Springer-Verlag New York, Inc., 1975, p. 229 
\bibitem[Renzini(1977)]{renzini1977} Renzini, A. 1977, in Advanced Stages in Stellar Evolution, ed. I. Iben, Jr., A. Renzini, \& D. N. Schramm ( Heidelberg: Springer)
\bibitem[Renzini(2008)]{renzini2008}Renzini, A.\ 2008, \mnras, 391, 354 
\bibitem[Renzini(2013)]{renzini2013} Renzini, A.\ 2013, arXiv:1302.0329 
\bibitem[Rutledge et al.(1997)]{rutledge1997} Rutledge, G.~A., Hesser, J.~E., \& Stetson, P.~B.\ 1997, \pasp, 109, 907 
\bibitem[Salaris et al.(2008)]{salaris2008} Salaris, M., Cassisi, 
S., \& Pietrinferni, A.\ 2008, \apjl, 678, L25 
\bibitem[Schaerer \& Charbonnel(2011)]{schaerer2011} Schaerer, D., \& Charbonnel, C.\ 2011, \mnras, 413, 2297 
\bibitem[Starkenburg et al.(2010)]{starkenburg2010} Starkenburg, E., Hill, V., Tolstoy, E., et al.\ 2010, \aap, 513, A34 
\bibitem[Strader et al.(2003)]{strader2003} Strader, J., Brodie, J.~P., Forbes, D.~A., Beasley, M.~A., 
\& Huchra, J.~P.\ 2003, \aj, 125, 1291 
\bibitem[Sweigart \& Gross(1976)]{sweigart-gross1976} Sweigart, A.~V., \& Gross, P.~G.\ 1976, \apjs, 32, 367 
\bibitem[Tolstoy et al.(2001)]{tolstoy2001} Tolstoy, E., Irwin, M.~J., Cole, A.~A., et al.\ 2001, \mnras, 327, 918 
\bibitem[Ventura et al.(2013)]{ventura2013} Ventura, P., Di Criscienzo, M., Carini, R., \& D'Antona, F.\ 2013, \mnras, 431, 3642 
\bibitem[Vesperini \& Heggie(1997)]{vesperini1997} Vesperini, E.,  Heggie, D.C., 1997, \mnras, 289, 898 
\bibitem[Vesperini et al.(2009)]{vesperini2009} Vesperini, E.,  McMillan, S.~L.~W., \& Portegies Zwart, S.\ 2009, \apj, 698, 615 
\bibitem[Vesperini et al.(2010)]{vesperini2010} Vesperini, E., McMillan, S.~L.~W., D'Antona, F., \& D'Ercole, A.\ 2010, \apjl, 718, L112 
\bibitem[Vesperini et al.(2013)]{vesperini2013} Vesperini, E., McMillan, S.~L.~W., D'Antona, F., \& D'Ercole, A.\ 2013, \mnras, 429, 1913 
\end{thebibliography}
\end{document}